\documentclass{aa}

\usepackage{graphicx}
\usepackage{amsmath}
\usepackage{amssymb}
\usepackage{natbib}
\usepackage{url}
\usepackage{epsfig}

\usepackage{txfonts}
\begin{document}

\title{The Building the Bridge Survey for $z=3$ Ly$\alpha$ emitting galaxies. II. Completion of the survey
\thanks{
Based on observations collected at the European Organisation for Astronomical
Research in the Southern Hemisphere, Chile, under programs 67.A-0033,
267.A-5704, 69.A-0380, 70.A-0048, and 072.A-0073.
}}

\author{L.F. Grove\inst{1}
\and J.~P.~U. Fynbo\inst{1} 
\and C. Ledoux\inst{2}
\and M. Limousin\inst{1,3}
\and P. M{\o}ller\inst{4}
\and K. Nilsson\inst{5}
\and B. Thomsen\inst{6}
}

\offprints{L.F. Grove,
lisbeth@dark-cosmology.dk}

\institute{ 
Dark Cosmology Centre, Niels Bohr Institute, University of
Copenhagen, Juliane Maries Vej 30, 2100 Copenhagen, Denmark 
\and
European Southern Observatory, Alonso de C{\'o}rdova 3107, Casilla
19001, Vitacura, Santiago 19, Chile
\and
Laboratoire d'Astrophysique de Toulouse-Tarbes, Universit{\'e} de Toulouse, CNRS,
57 avenue d'Azereix, 65000 Tarbes, France
\and
European Southern Observatory, Karl-Schwarzschild-Strasse 2, D-85748 Garching,
Germany
\and
Max-Planck-Institut f{\"u}r Astronomie, K{\"o}nigstuhl 17, 69117 Heidelberg, Germany
\and 
Institute of Physics and Astronomy, Aarhus University, Ny Munkegade, DK-8000 Aarhus C
}

\date{Received -; Accepted -}

\abstract{
%Context
We have substantial information about kinematics and abundances of
galaxies at $z\approx3$ studied in absorption against the light of
background QSOs. At the same time we have studied 1000s of galaxies
detected in emission mainly through the Lyman-break selection
technique. However, we know very little about how to make the
connection between the two data sets.}  
{%Aims 
We aim at bridging the gap between absorption selected and
emission selected galaxies at $z\approx3$ by probing the faint end of
the luminosity function of star-forming galaxies at $z\approx3$.}
{%Methods 
Narrow-band surveys for Lyman-$\alpha$ (Ly$\alpha$) emitters have proven
to be an efficient probe of faint, star-forming galaxies in the high
redshift universe. We have performed narrow-band imaging in three
fields with intervening QSO absorbers (a damped Ly$\alpha$
absorber and two Lyman-limit systems) using the VLT. We target
Ly$\alpha$ at redshifts 2.85, 3.15 and 3.20.}  
{%Results 
We find a consistent surface density of about 10
Ly$\alpha$-emitters per square arcmin per unit redshift in all three
fields down to our detection limit of about $3\times10^{41}
\rm{ergs\; s^{-1}}$. The luminosity function is consistent with what has been
found by other surveys at similar redshifts. About 85\% of the sources
are fainter than the canonical limit of $R=25.5$ for most Lyman-break
galaxy surveys. In none of the three fields do we detect the emission
counterparts of the QSO absorbers. In particular we do not detect the
counterpart of the $z=2.85$ damped Ly$\alpha$ absorber towards
Q2138$-$4427. This implies that the DLA galaxy is either not a
Ly$\alpha$ emitter or fainter than our flux limit.  } 
{%Conclusion 
Narrow-band surveys for Ly$\alpha$ emitters are
excellent to probe the faint end of the luminosity function at
$z\approx3$. There is a very high surface density of this class of
objects. Yet, we only detect galaxies with Ly$\alpha$ in emission
and hence the density of galaxies with similar broad band magnitudes
will be substantially higher. This is consistent with a very
steep slope of the faint end of the luminosity function as has been
inferred by other studies. This faint population of galaxies is playing
a central role in the early Universe. There is evidence that this population
is dominating the intergrated star-formation activity, responsible for 
the bulk of the ionizing photons at $z\gtrsim3$ and likely also responsible
for the bulk of the enrichment of the intergalactic medium.}  

\maketitle

\section{Introduction}
\label{sec:intro}

Strong arguments
\citep{fynbo1999,haehnelt2000,schaye2001,rauch2008,barnes2008}
indicate that there is very little overlap between emission selected
galaxies \citep[primarily Lyman-Break Galaxies, LBGs,][]{steidel2003}
and absorption selected galaxies \citep[primarily the Damped
Lyman-$\alpha$ Absorbers, DLAs][]{wolfe2005}.  The simple
reason for this is that LBG samples are continuum flux limited and
that the current flux limit of R$\approx$25.5 is not deep enough to
reach the level of typical absorption selected galaxies. This is
unfortunate as we then know little about how to combine the detailed
information on abundances and kinematics inferred from observations of
DLAs with the information about colours and luminosities of high-$z$
galaxies detected in emission.

In 2000 we started a survey aiming at bridging the gap between
emission and absorption selected galaxies.  The goal of the survey was
to detect faint $z\approx3$ galaxies using narrow-band imaging
selection of Lyman-$\alpha$ (Ly$\alpha$) emitters (LAEs) and in this
way bridge the gap between the DLAs and the LBGs. During the 1990ies
and early 2000s it was established that LAEs can be used to select
high-z galaxies \citep[e.g.][]{moeller1993} and that this method
easily traces significantly deeper into the luminosity function than
what is possible with spectroscopic samples of LBGs
\citep[e.g.][]{cowie1998,fynbo2001}.  In our survey we targeted the
fields of QSOs with intervening DLAs primarily to be able also to
search for the galaxy counterparts of the DLAs, but also to anchor the
fields to known structures at the targeted redshifts.  The first
paper of the survey was published by \cite{fynbo2003} (hereafter
Paper~I), where we presented the results from two of the three
targeted fields. Since then the study of LAEs has progressed
substantially, mainly on two fronts. First, large samples covering a
range of redshifts have been collected using wide-field imagers both
on 4m and 8m class telescopes
\citep[e.g.][]{gronwall2007,ouchi2008,nilsson2008}. Second, more
detailed studies of the properties have been carried through, most
notably based on LAEs in the GOODS fields
\citep[e.g.][]{nilsson2007,gronwall2007,pentericci2008}.

In this final paper from our ''Building the bridge'' survey we first
present the results from the third field, namely the sample of LAEs
identified in the field of the quasar BRI\,1202$-$0725. Second, we
combine the results of the entire survey covering three fields to
derive the luminosity function of the LAEs at $z\sim3$. We then
discuss our results in the light of the substantial progress that has
been made by a number of groups over the last few years on the study
of LAEs at $z\gtrsim2$.

Throughout this paper, we assume a cosmology with $H_0=70$ km s$^{-1}$
Mpc$^{-1}$, $\Omega _{\rm m}=0.3$ and $\Omega _\Lambda=0.7$.  We also
throughout use magnitudes on the AB system \citep{oke1974}.

\section{The field of BRI\,1202$-$0725}
\label{sec:q1202}

\subsection{Imaging}
\label{sec:imaging}

\begin{figure}
\resizebox{\columnwidth}{!}{\includegraphics{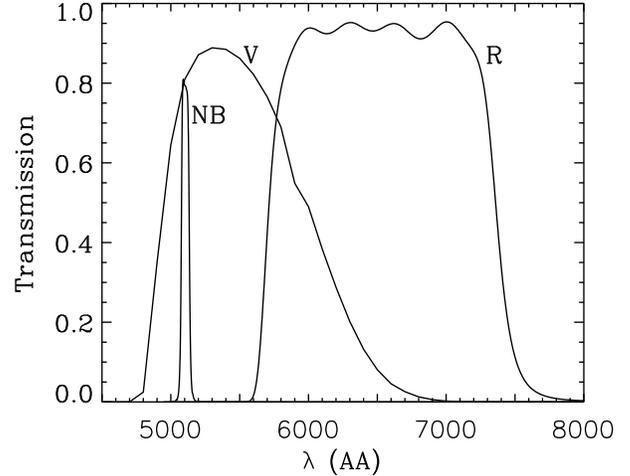}}
\caption{Transmission curves of the narrow- and broad-band filters
used for the observations of BRI\,1202$-$0725.}
\label{fig:filtercurves}
\end{figure}

The field of BRI\,1202$-$0725 ($z_{\rm qso}=4.70$) was included in this survey due to the presence of
a Lyman-limit system along the line-of-sight towards the quasar at a redshift
of $z=3.2$ \citep{storrie-lombardi1996}. This field was observed in service
mode at the VLT 8.2 m telescope, unit Yepun, during the nights January 30
through February 6 2003 using the FORS2 instrument. The wavelength of the
Ly$\alpha$ transition at the redshift of $z=3.20$ is $\lambda=5105{\AA}$, which
corresponds to the central wavelength of the 61\AA \ wide [OIII] VLT filter.
The field was observed in this narrow-band filter (NB) and in the Bessel $V$
and Special $R$ filters. The transmission curves of the three filters are shown
in Fig.~\ref{fig:filtercurves}. The integration times in the $V$ and $R$
filters were set by the criterion that the broad on-band $V$ imaging should
reach about half a magnitude deeper than the narrow-band imaging so as to get a
reliable selection of objects with {\it excess} emission in the narrow-band
filter. For the broad off-band $R$ imaging, we aimed at reaching the $5\sigma$
significance level at one magnitude deeper than the spectroscopic limit of
$R$(AB)=25.5 for LBGs (i.e.  aiming at $R$(AB)=26.5 at the $5\sigma$
significance level). The total integration times, the seeing (FWHM) of the
combined images and the 5$\sigma$ detection limits for a 3\arcsec diameter
aperture are given for each filter in Table~\ref{tab:imglog}. For comparison
with previous work we also list the corresponding numbers for the fields of
BRI\,1346$-$0322 and Q\,2138$-$4427 presented in Paper~I\footnote{Note that due
to an error in the photometric zeropoints used in that paper the listed broad
band detection limits are slightly different from the original ones (by about
0.15 mag). This is not affecting the conclusions of that paper.}.

\begin{table}
\caption{Log of imaging observations with FORS1 and 2 of the three
surveyed fields. The $5\sigma$ detection limits are computed for a
3\arcsec diameter aperture. The data for BRI\,1346$-$0322 and
Q\,2138$-$4427 are reproduced from Paper~I.}
\label{tab:imglog}
\begin{minipage}{\columnwidth}
\begin{tabular}{lllll}
\hline
filter & $z_{\rm qso}$, $z_{\rm abs}$ & total exp. & PSF fwhm & 5$\sigma$\\
       & & time (hr)  & ($\arcsec$)     & limit    \\
\hline
BRI\,1202$-$0725 &4.70, 3.20\\
$N$ & & 8.3 & 1.02 & 25.6\\
$V$\footnote{Due to the higher target redshift of this field the wavelength of the Ly$\alpha$-line was matched by the $V$-band rather than $B$.}
  & & 2.9 & 0.96 & 26.6 \\
$R$ & & 1.6 & 0.95 & 26.1\\
\hline
BRI\,1346$-$0322 & 3.99, 3.15 \\
$N$ & &  8.9    & 0.93  & 25.6\\
$B$      & &  2.5    & 1.02  & 26.6 \\
$R$      & &  1.7    & 0.94  & 26.0 \\
\hline
Q\,2138$-$4427 & 3.17, 2.85 \\
$N$ & &  10.0   &  0.96  & 26.5\\
$B$      & &  2.5    &  1.04  & 26.9 \\
$R$      & &  1.7    &  0.93  & 26.3 \\
\hline
\label{journal}
\end{tabular}
\end{minipage}
\end{table}

The images were reduced (de-biased and corrected for CCD pixel-to-pixel
variations) using the FORS pipeline \citep{grosboel1999}. The
individual reduced images in each filter were combined using a code that
optimizes the Signal-to-Noise (S/N) ratio for faint, sky-dominated sources
(see M\o ller \& Warren 1993 for details on this code).

The broad-band images were calibrated as part of the FORS calibration
plan via observations of Landolt standard stars (Landolt 1992). We
transformed the zero-points to the AB system using the relations given
by \cite{fukugita1995}: $V$(AB)=$V$$-0.02$ and $R$(AB)=$R$$+0.17$. For the
calibration of the narrow-band images, we used observations of the
spectrophotometric standard stars EG274 and GD71.

\subsection{LAE candidate selection}
\label{sec:LAE_selection}

The selection of LAE candidates is based on the ``narrow minus on-band
broad'' versus ``narrow minus off-band broad'' colour/colour plot
technique
\citep{moeller1993,fynbo1999,fynbo2000c,fynbo2002,fynbo2003}. The
object identification and photometry was carried out in SExtractor
\citep{bertin1996} using the dual-image mode having a detection image
and measuring the photometric properties on the individual images. For
the detection we used a weighted sum of the $V$-band (20\%) and
narrow-band (80\%) images to secure an optimal detection of objects
with excess in the narrow filter. Before object detection we convolved
the detection image with a Gaussian filter function having a FWHM
equal to that of point sources. We used a detection threshold of 1.1
times the background sky-noise and a minimum area of 5 connected
pixels in the filtered image. For total magnitudes we use the
SExtractor MAG\_AUTO and to compute colours we use the isophotal
magnitudes (MAG\_ISO).  In our final catalogue, we include only
objects with total S/N$>$5 in the isophotal aperture in either the
narrow- or the $V$-band images. In total, we detect 3202 such objects
within the 400$\times$400 arcsec$^2$ field around BRI\,1202$-$0725. The
error bars on the colour indices are derived using the maximum
likelihood method of \cite{fynbo2002}. For the selection of LAE
candidates we restricted the sample to S/N$>$5 in the narrow-band
filter as described below.

\begin{figure*}
\begin{center}
\resizebox{0.32\textwidth}{!}{\includegraphics{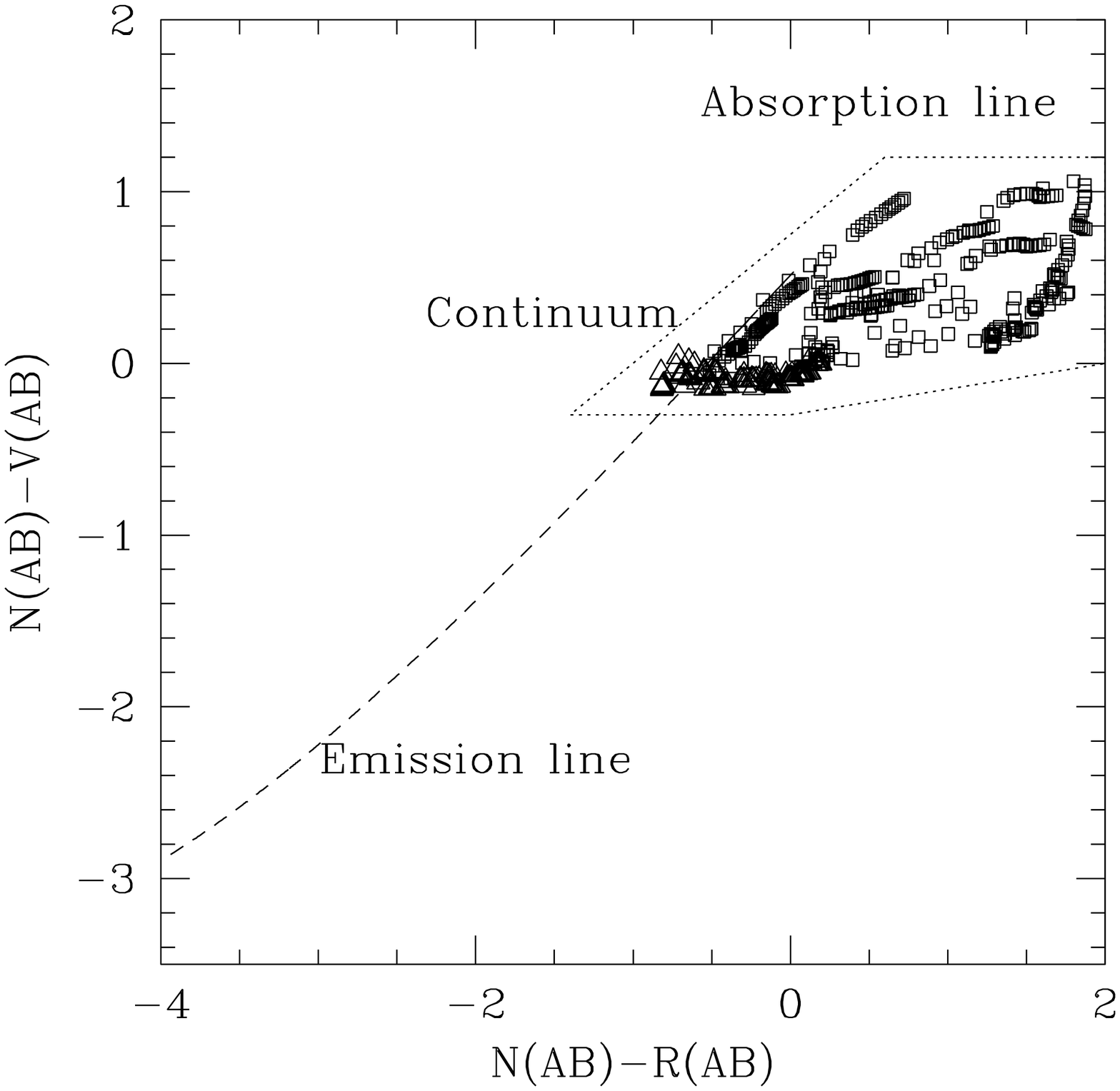}}
\resizebox{0.32\textwidth}{!}{\includegraphics{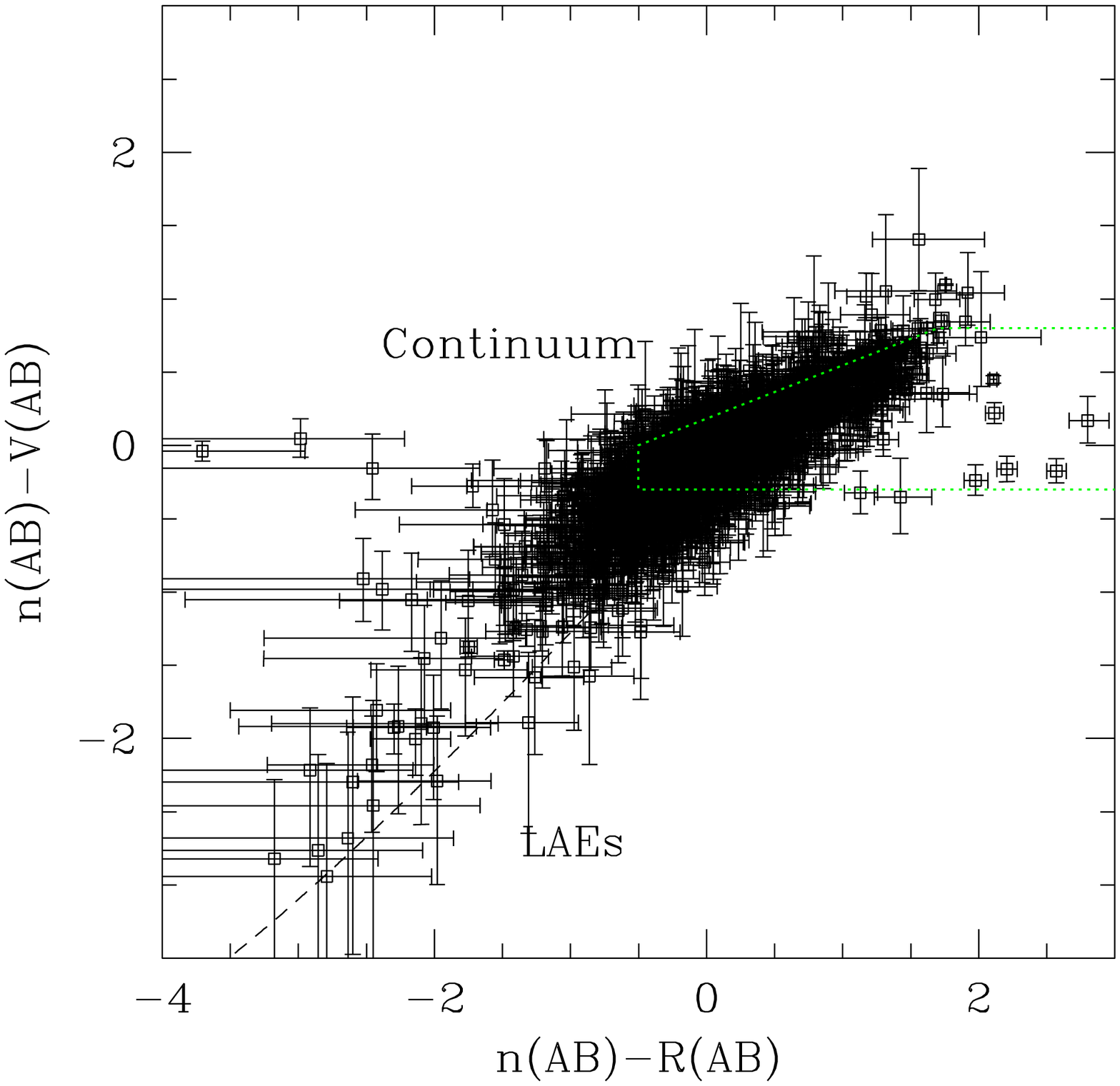}}
\resizebox{0.32\textwidth}{!}{\includegraphics{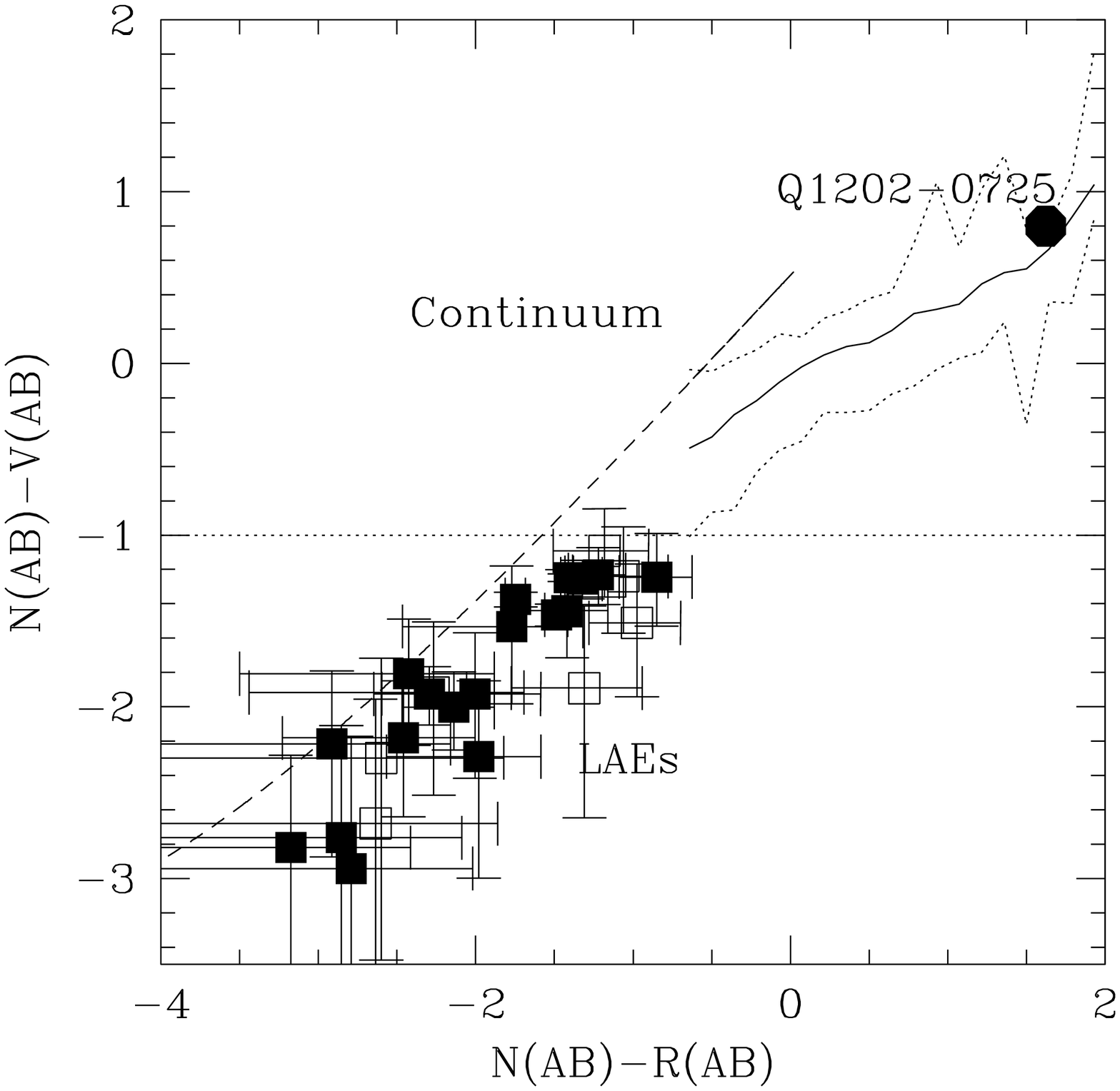}}
\end{center}
\caption{ {\it Left panel:} Colour-colour diagram for simulated
galaxies based on Bruzual \& Charlot 1995 galaxy SEDs
\citep{bruzual2003}. The open squares are 0$<$z$<$1.5 galaxies with
ages ranging from a few to 15 Gyr and the open triangles are
1.5$<$z$<$3.0 galaxies with ages ranging from a few Myr to 1Gyr. The
dotted box encloses the simulated galaxy colours. The dashed line
indicates colours of objects with a particular broad-band colour and
with an SED in the narrow-band filter ranging from an absorption line
in the upper right part to a strong emission line in the lower left
part of the diagram. {\it Middle panel:} Colour-colour diagrams for
all objects detected at S/N$>$5 in either the narrow band or the $V$
band in the BRI\,1202$-$0725. In the lower left part of the diagram
there is a large number of objects with excess emission in the narrow
filter. Candidate LAEs are those objects with a 1$\sigma$ upper limit
on $N$(AB)$-V$(\rm{AB}) below the 98\% percentile.  {\it Right panel:}
The colours of the QSOs and of the LAE candidates are
indicated. Candidates confirmed to be emission line objects based on
our spectroscopic observations described in Sect.~\ref{sec:MOS} are
shown by filled symbols. Candidates that cannot be confirmed by the
present spectroscopy are indicated by open symbols. The solid line
indicates the median $N(\rm{AB})-V(\rm{AB})$ colour for all detected
objects as a function of $N(\rm{AB})-R(\rm{AB})$ and the dotted lines
indicate the 2\% and 98\% percentiles in the $N(\rm{AB})-V(\rm{AB})$
colour. The horisontal dotted line denote our selection of
$N(\rm{AB})-V(\rm{AB})<-1$.  }
\label{fig:colcol}
\end{figure*}

In Fig.~\ref{fig:colcol} we show the colour-colour diagram used for
the final selection of LAE candidates. In order to constrain where
objects with no special spectral features in the narrow filter are in
the diagram, we calculated colours based on synthetic galaxy SEDs
taken from the Bruzual \& Charlot 1995 models \citep{bruzual2003}.  We
have used simple single-burst models with ages ranging from a few Myr
to 15 Gyr and with redshifts from 0 to 1.5 (open squares in
Fig.~\ref{fig:colcol}) and models with ages ranging from a few Myr to
1 Gyr with redshifts from 1.5 to 3.0 (open triangles). For the colours
of high-redshift galaxies, we included the effect of Ly$\alpha$
blanketing \citep{moeller1990,madau1995}, but for none of the models
the effects of dust are included. Fig.~\ref{fig:colcol} shows the
N(AB)$-V$(AB) versus N(AB)$-$$R$(AB) colour diagram for the simulated
galaxy colours (left panel) and for the observed sources in the target
field (middle and right panels). The dashed line indicates where
objects with a particular broad-band colour (corresponding to a 100
Myr old galaxy at $z=3.20$ and either absorption (upper right) or
emission (lower left) in the narrow filter will fall.

In the middle panel, we show the colour-colour diagram for all of the
objects detected in the field. The large, dense group of points
correspond to the normal field galaxy population without special
properties in the narrow-band filter, in the following refered to as
continuum objects. Unfortunately, the observed distribution of the
continuum objects is oriented along the same direction as that
expected for emission line objects (indicated with the dashed
line). This is due to the fact that the central wavelength of the
narrow-band filter is bluewards of the central wavelength of the
on-band broad filter ($V$). For this reason we for this field
choose a rather conservative criterion for selection of candidates,
namely $N(\rm{AB})-V(\rm{AB})<-1$.  After weeding out a few spurious
sources (related to bright stars) we detect 25 such
objects in the BRI\,1202$-$0725 field which we consider as LAE
candidates in the following. The colours of the candidates are shown
in the right panel of the figure.

\begin{figure}
\resizebox{\columnwidth}{!}{\includegraphics{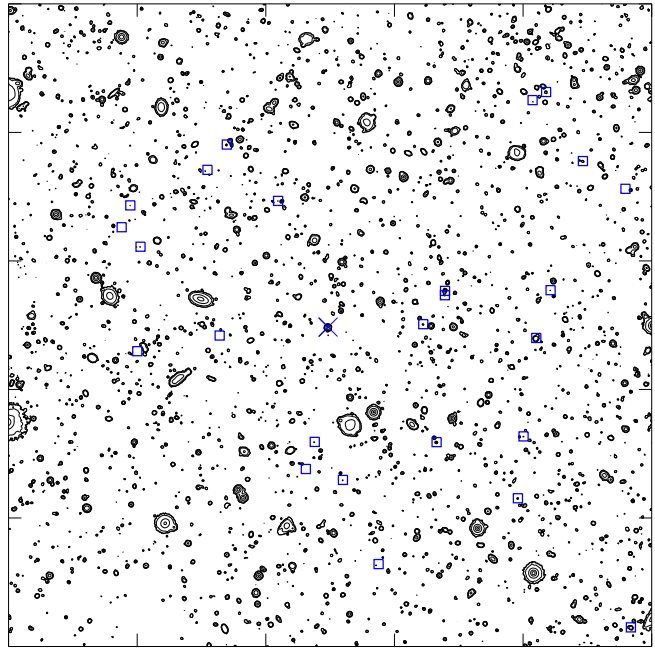}}
\caption{The 400$\times$400 arcsec$^2$ field surrounding the QSO
BRI\,1202$-$0725 as observed in the narrow-band filter. North is up and
East is to the left.  The QSO is identified by a ``$\times$'' at the
field centre and the positions of candidate LAEs are shown with
boxes.}
\label{fig:fieldmap}
\end{figure}

A contour image of the combined narrow-band image of the
400$\times$400 arcsec$^2$ field surrounding the QSO BRI\,1202$-$0725 is
shown in Fig.~\ref{fig:fieldmap}. The QSO is identified by a
``$\times$'' at the field centre and the positions of selected LAE
candidates are shown with boxes.

\begin{figure}
\begin{center}
\resizebox{\columnwidth}{!}{\includegraphics{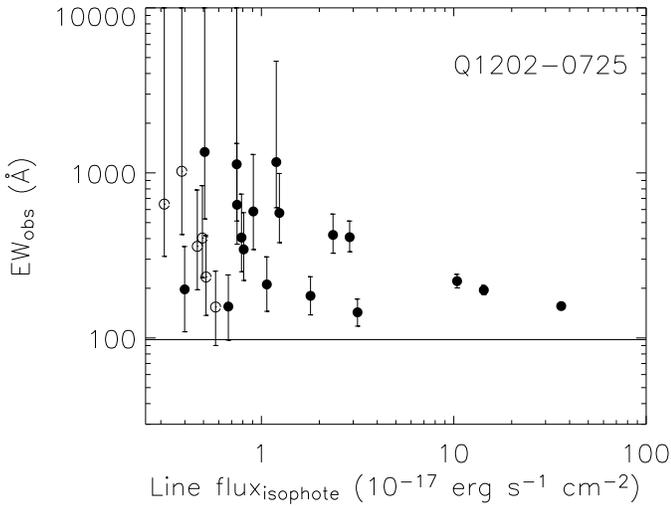}}
\caption{Our equivalent width selection criterion is illustrated by
plotting the line equivalent width against the Ly$\alpha$ flux in the
isophotal aperture for the LAE candidates. The continuous line shows
our N(AB)$-$$V$(AB) colour selection criterion converted into equivalent
width. Filled symbols indicate objects subsequently confirmed by
spectroscopy to be emission-line sources. Open circles indicate
objects that were observed, but not confirmed. The error bars on
EWs are derived as described in Fynbo et al.\ (2002).}
\label{fig:ew_select}
\end{center}
\end{figure}

In the selection of LAE candidates we apply a colour selection which
translates into the equivalent width (EW) selection criterion
illustrated in Fig.~\ref{fig:ew_select} given in observed
quantities. The selection criterion translates into candidates having
observed EW$\gtrsim100$\AA\ (or rest frame EW$\gtrsim25$\AA ).

\subsection{Multi-object spectroscopy}
\label{sec:MOS}

The above selection of LAE candidates only selects for excess emission
in the narrow-band filter. This is likely to originate from the
Ly$\alpha$ at $z\sim3.2$ \citep{fynbo2003}, however, interlopers caused
by other emission lines in galaxies at lower redshifts is also
possible. Therefore, follow-up spectroscopy is necessary for
confirming the Ly$\alpha$ origin of the excess emission.

Follow-up multi-object spectroscopy (MOS) was carried out in visitor
mode on March 21--23, 2004, with FORS2 installed at the VLT telescope,
unit Yepun. The mask preparation was done using the FORS Instrumental
Mask Simulator. The field of BRI\,1202$-$0725 was covered by 3 masks. It
was possible to fit all candidates into the slits. The spectra were
obtained with the G600B grism covering the wavelength range from
3600\AA \ to 6000\AA \ at a resolving power of 900. For possible
identification of other emission lines at the same redshift we also
obtained spectra with the G600R grism covering the wavelength range
from 5000\AA\ to 7500\AA . The detector pixels were binned 2$\times$2
for all observations through the masks. In
Table~\ref{tab:spec-journal} we give the main characteristics of the
spectroscopic observations.

\begin{table}
\begin{center}
\caption{Log of spectroscopic observations with FORS2.
}
\begin{tabular}{@{}lllcccc}
\hline
mask & Exp.time & Date   & Effective seeing \\
     &    (hr)  & (2004) &    ($\arcsec$)   \\
\hline
mask1202A & 6.3 & March 21-23  & 0.94 \\
mask1202B & 6.0 & March 21-23 &  0.76\\
mask1202C & 5.0 & March 21-23 &  0.96\\
\hline
\label{tab:spec-journal}
\end{tabular}
\end{center}
\end{table}

The MOS data were reduced and combined as described in
\cite{fynbo2001}. The accuracy in the wavelength calibration is about
$\pm$0.1 pixel for a spectral resolution of $R=900$, which translates
to $\Delta z=0.0002$. Average object extraction was performed within a
variable window size matching the spatial extension of the emission
line. Therefore, the flux should be conserved.

\begin{figure*}
\begin{center}
\resizebox{0.3\textwidth}{!}{\includegraphics{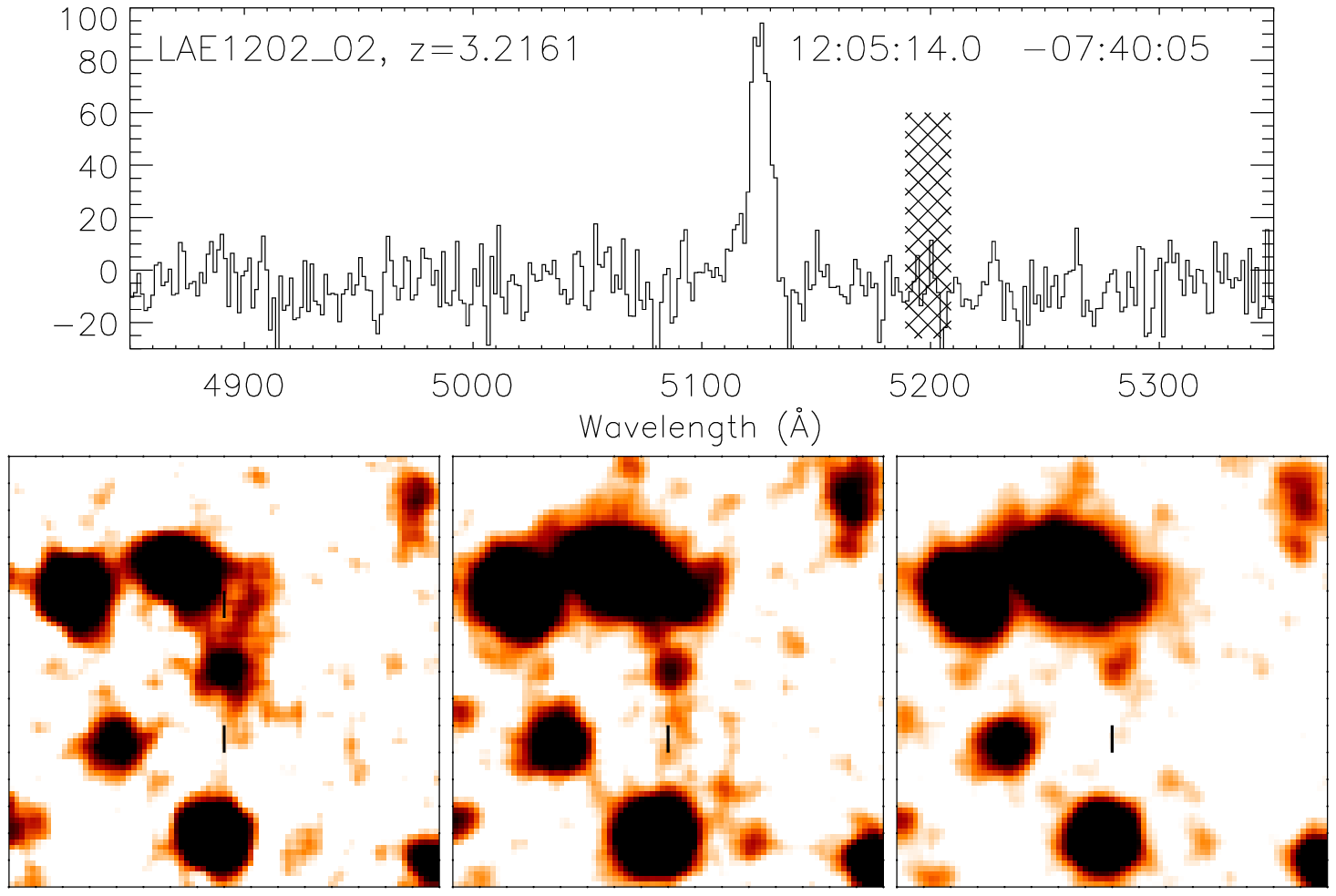}}
\resizebox{0.3\textwidth}{!}{\includegraphics{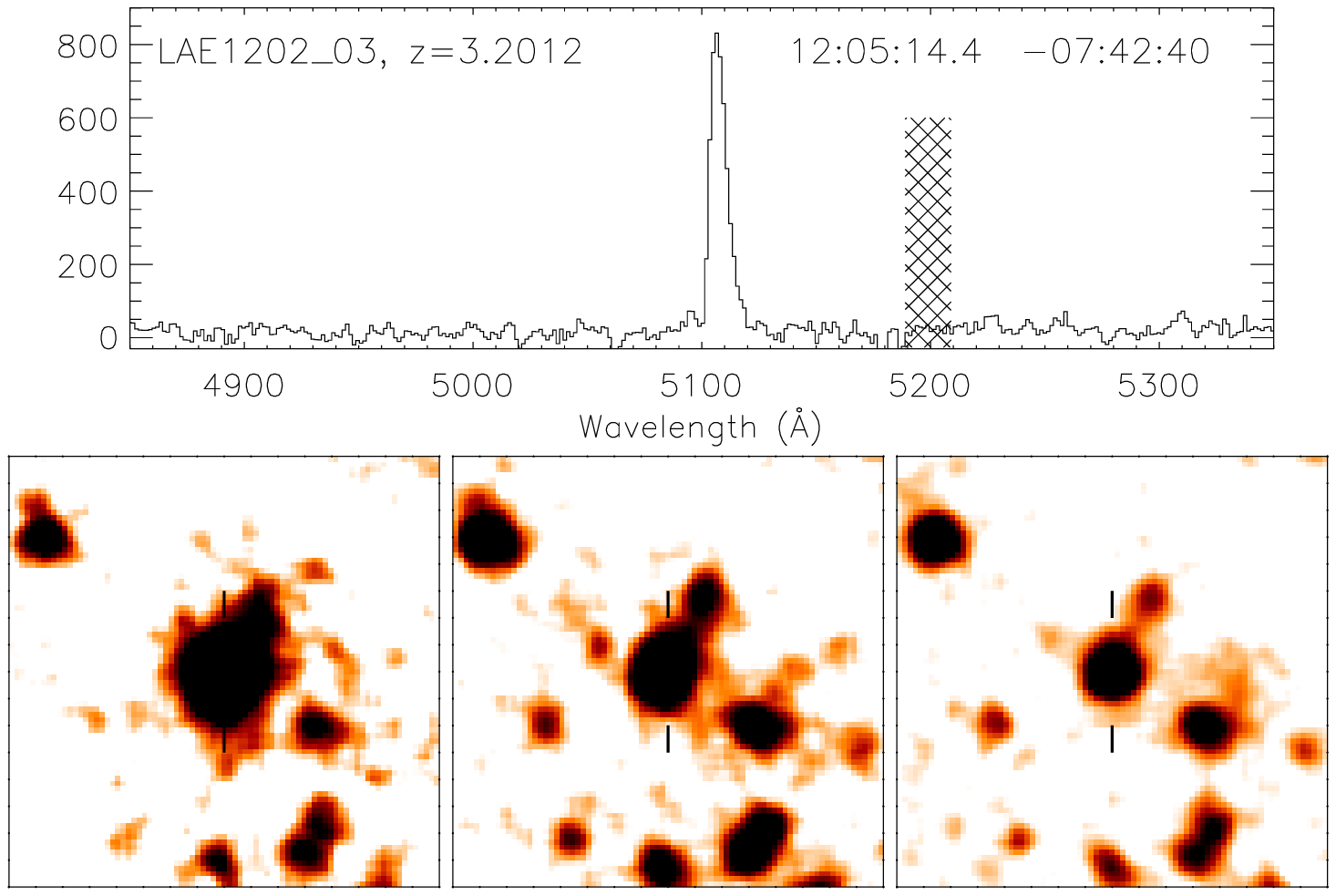}}
\resizebox{0.3\textwidth}{!}{\includegraphics{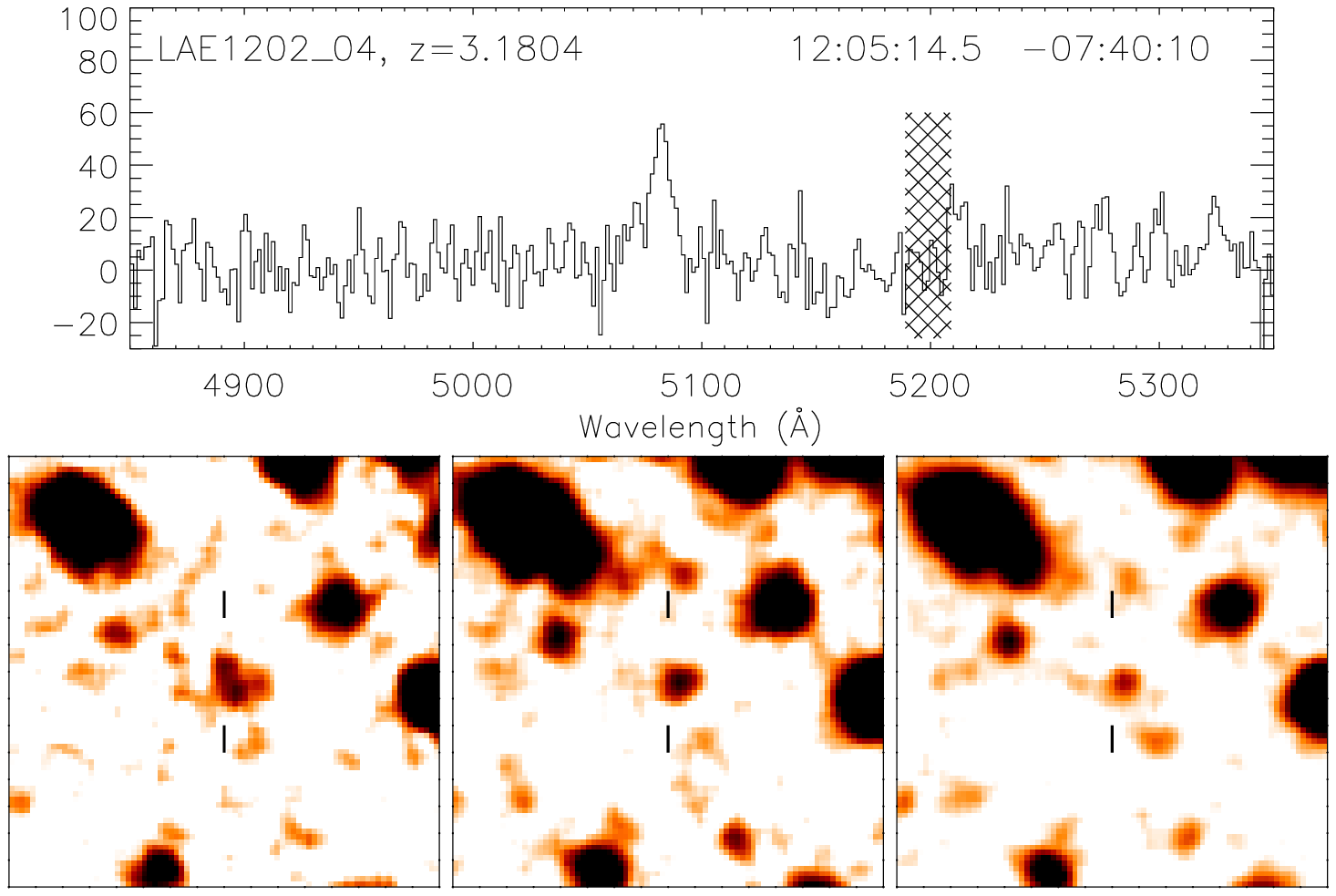}}
\resizebox{0.3\textwidth}{!}{\includegraphics{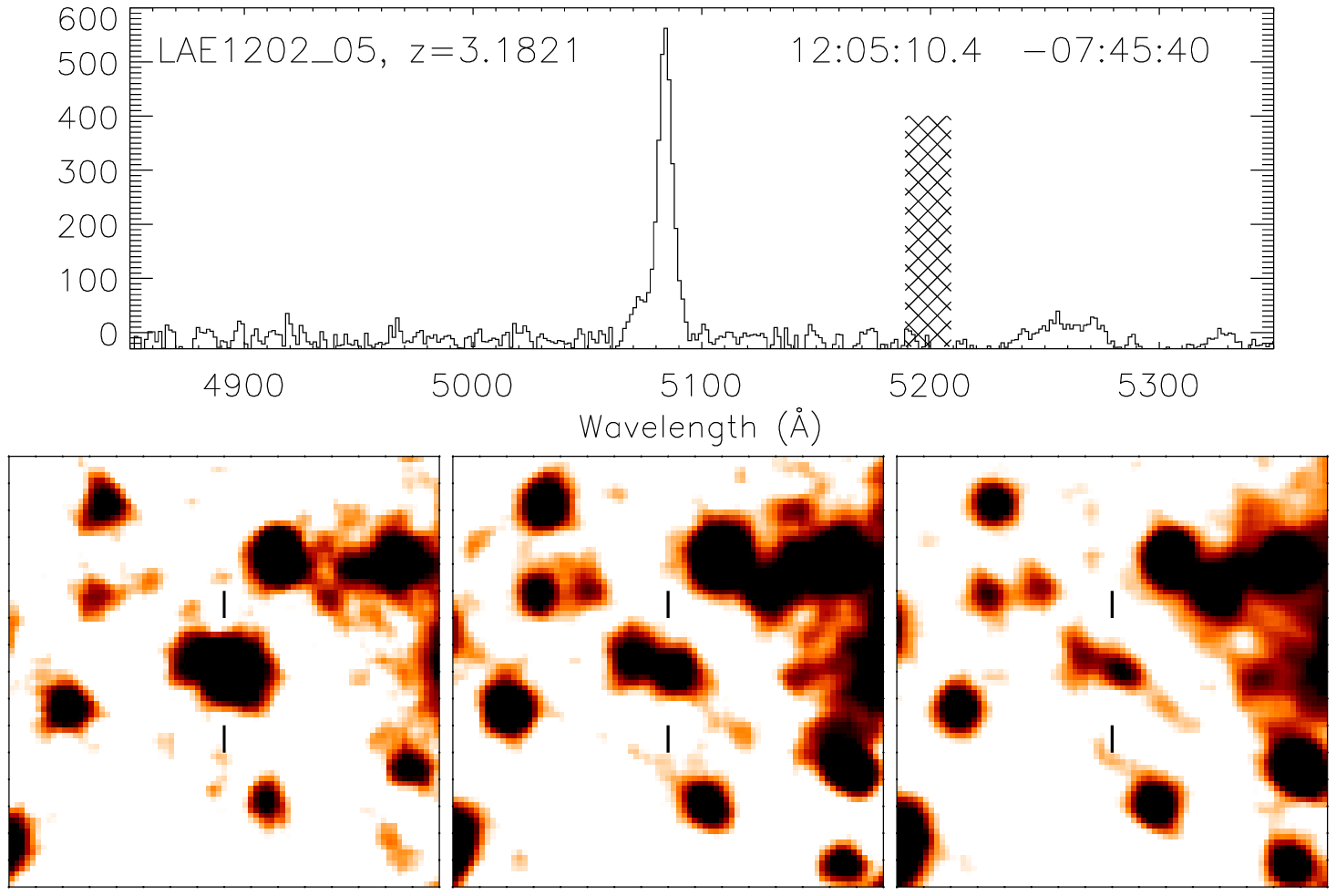}}
\resizebox{0.3\textwidth}{!}{\includegraphics{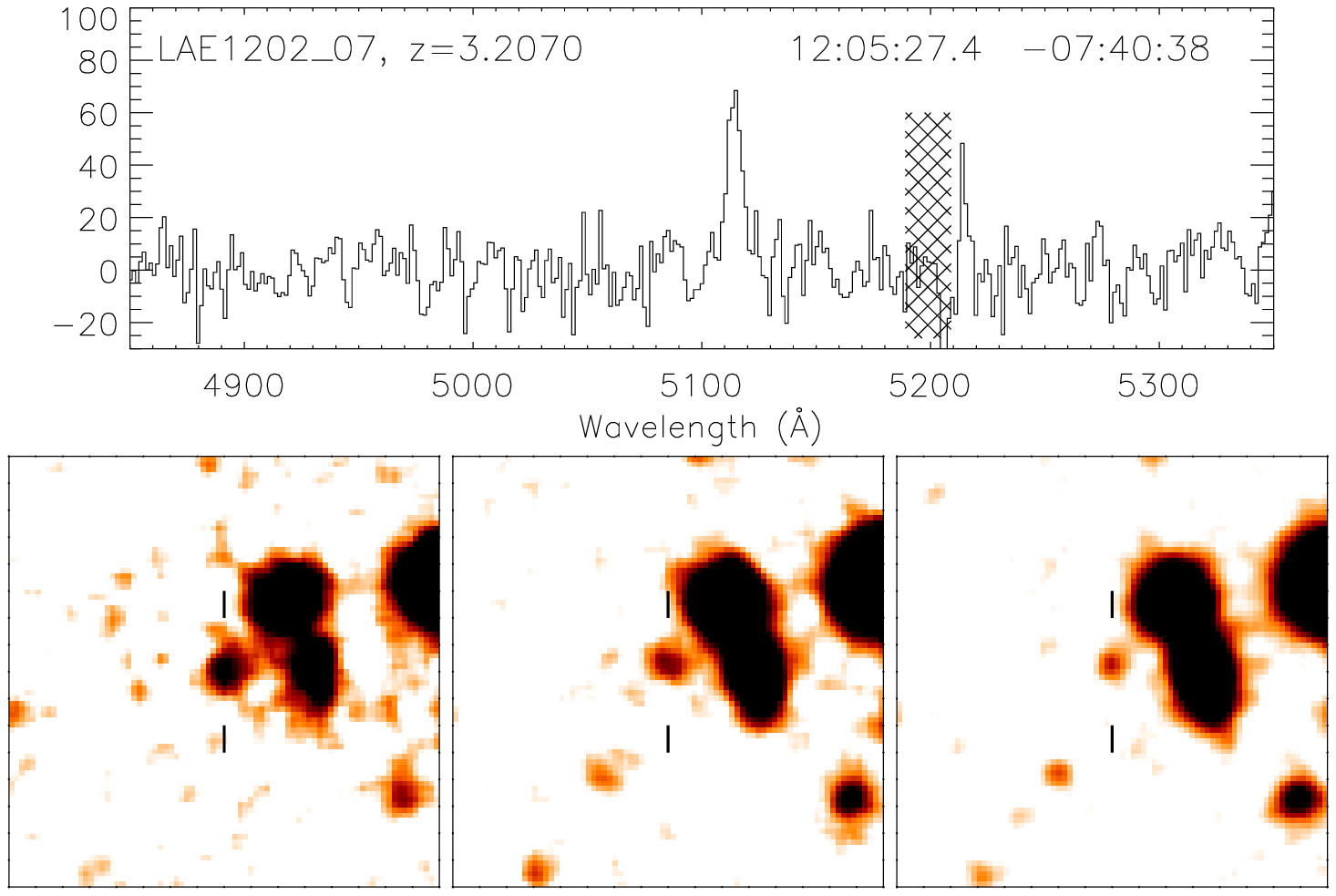}}
\resizebox{0.3\textwidth}{!}{\includegraphics{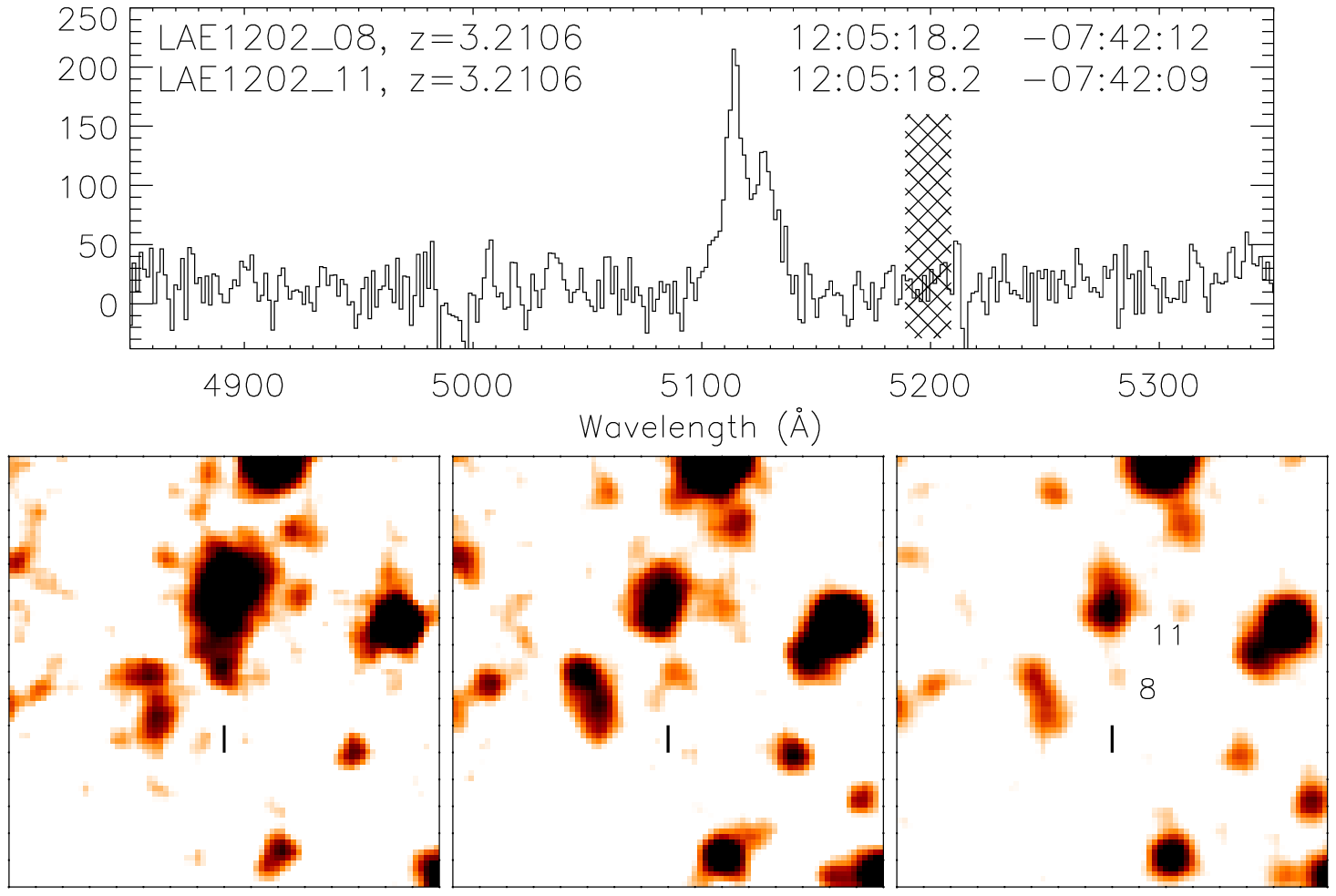}}
\resizebox{0.3\textwidth}{!}{\includegraphics{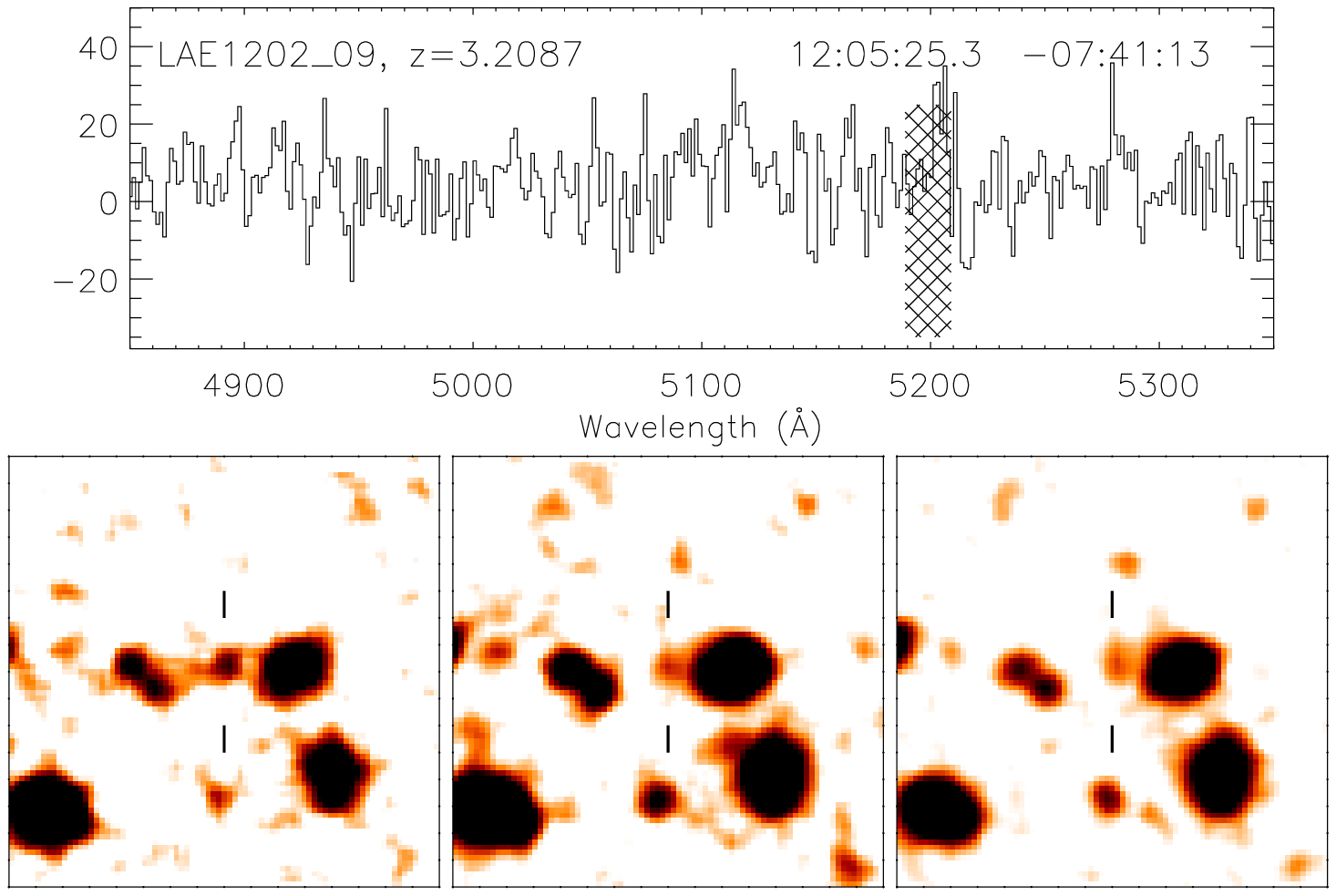}}
\resizebox{0.3\textwidth}{!}{\includegraphics{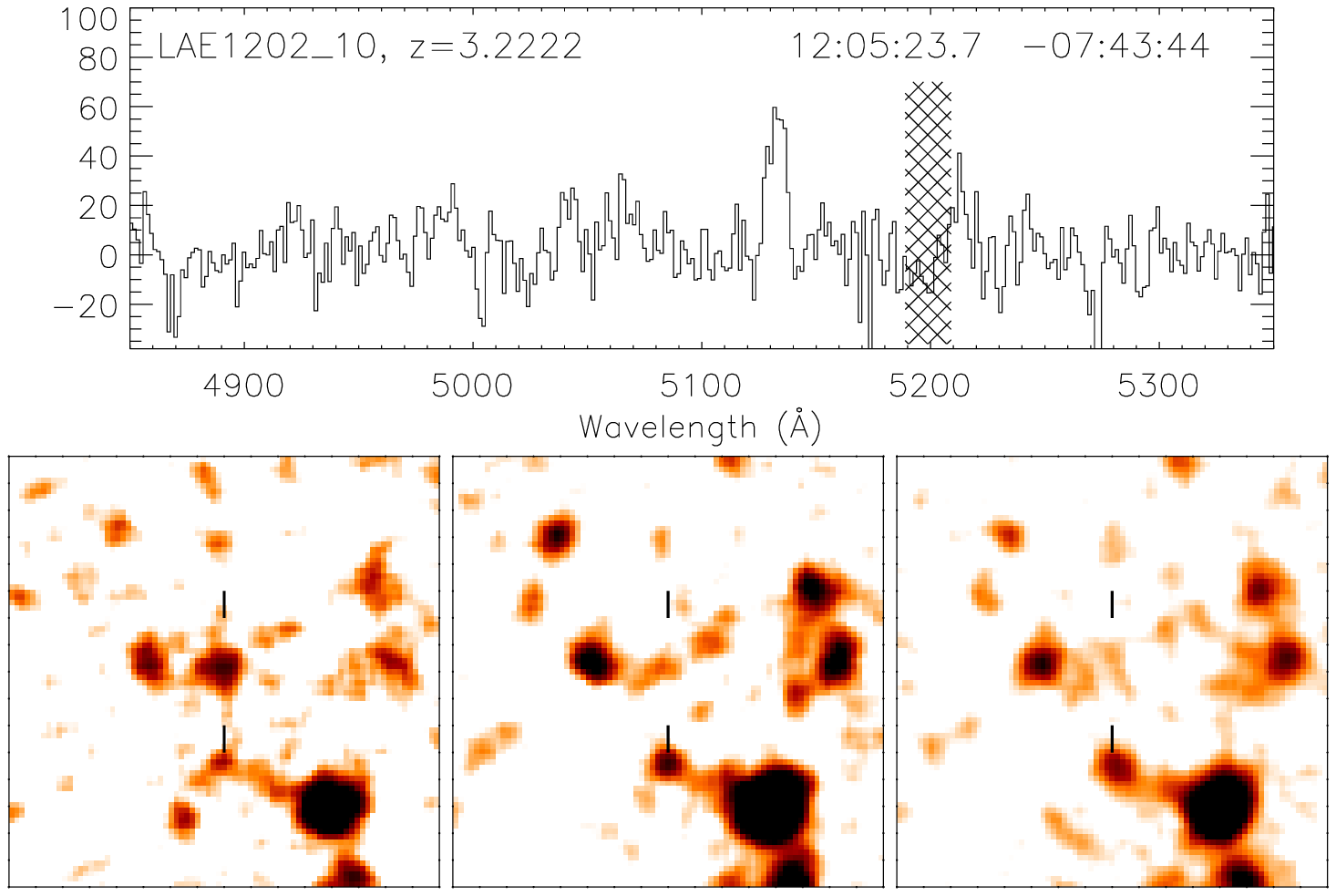}}
\resizebox{0.3\textwidth}{!}{\includegraphics{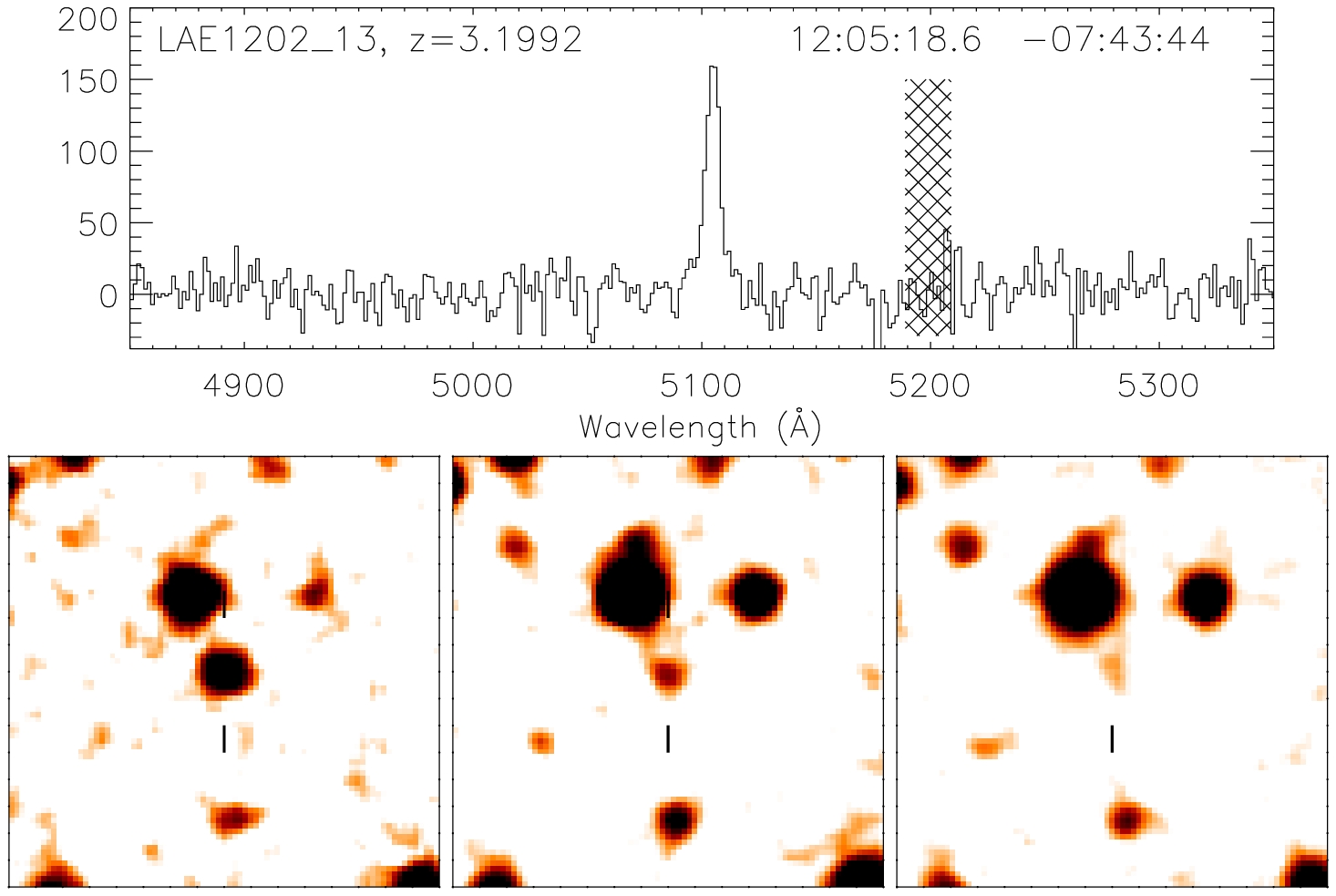}}
\resizebox{0.3\textwidth}{!}{\includegraphics{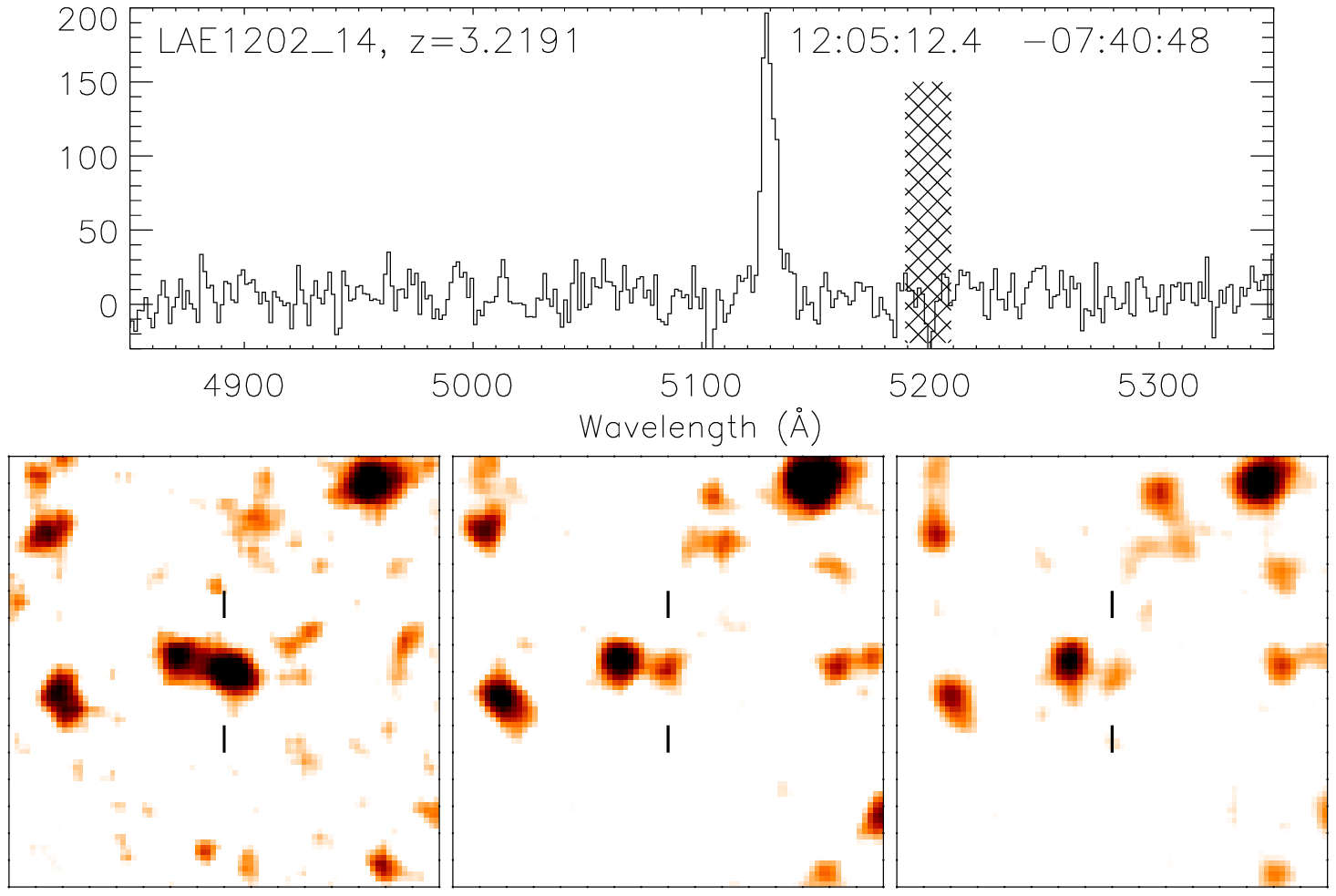}}
\resizebox{0.3\textwidth}{!}{\includegraphics{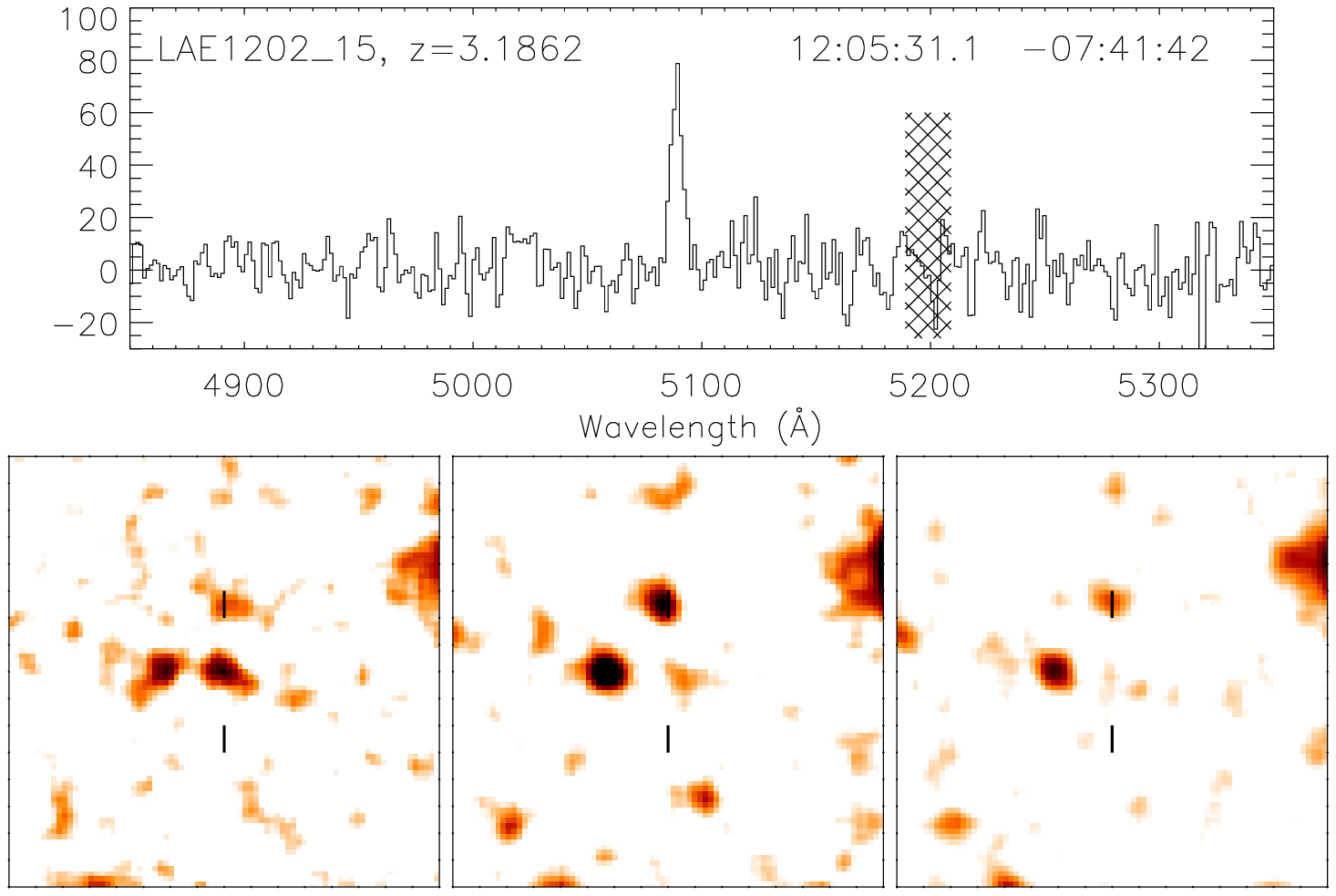}}
\resizebox{0.3\textwidth}{!}{\includegraphics{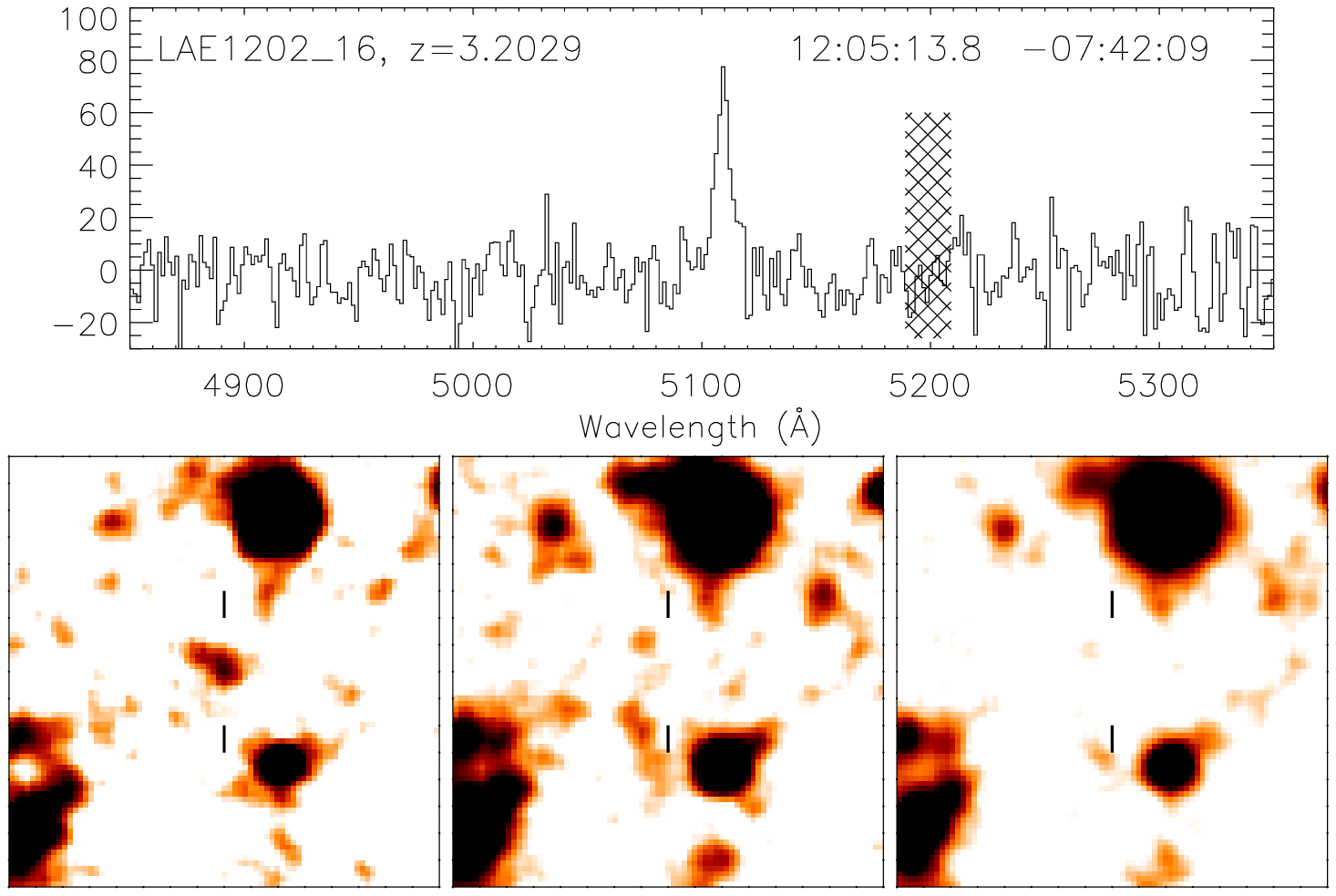}}
\resizebox{0.3\textwidth}{!}{\includegraphics{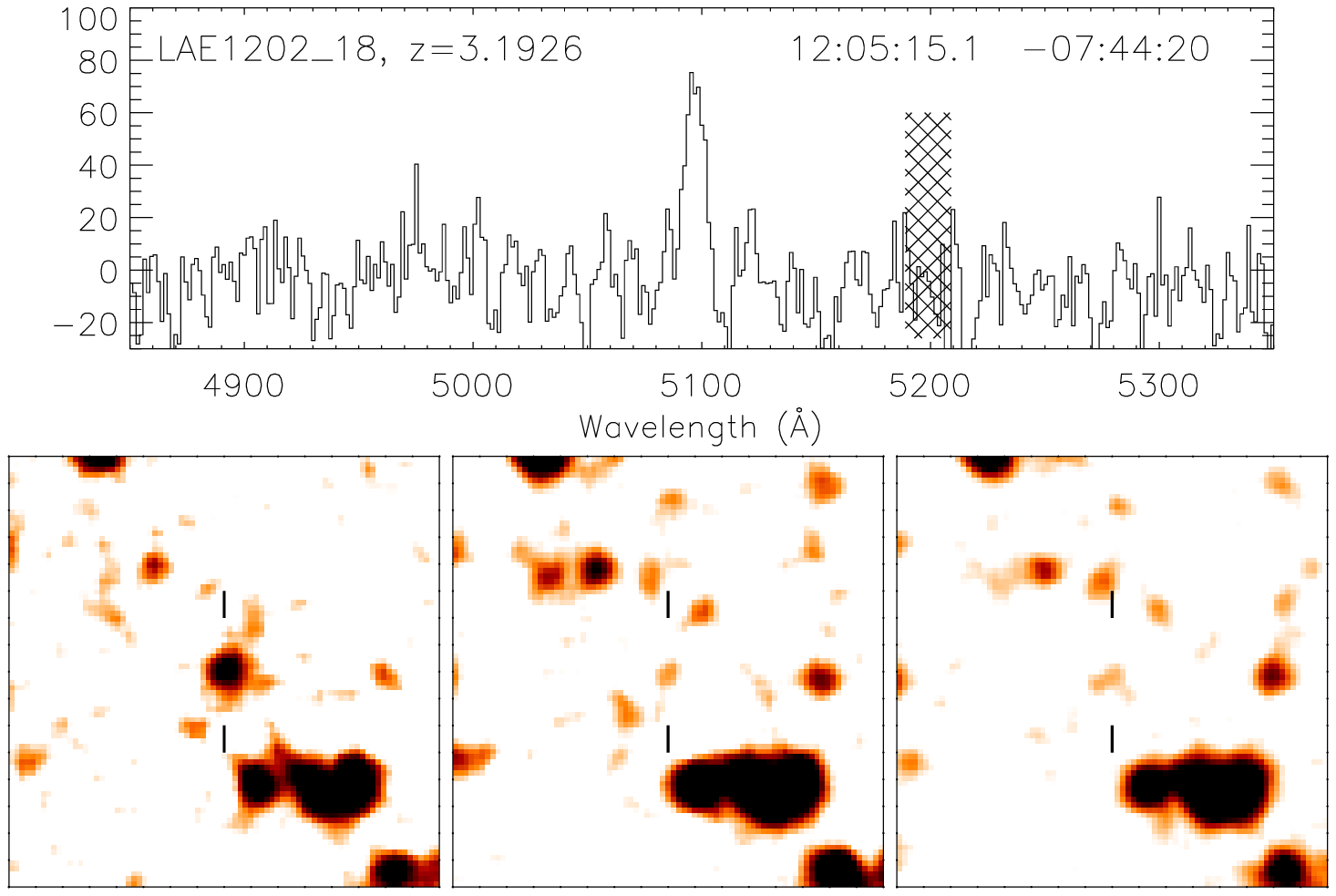}}
\resizebox{0.3\textwidth}{!}{\includegraphics{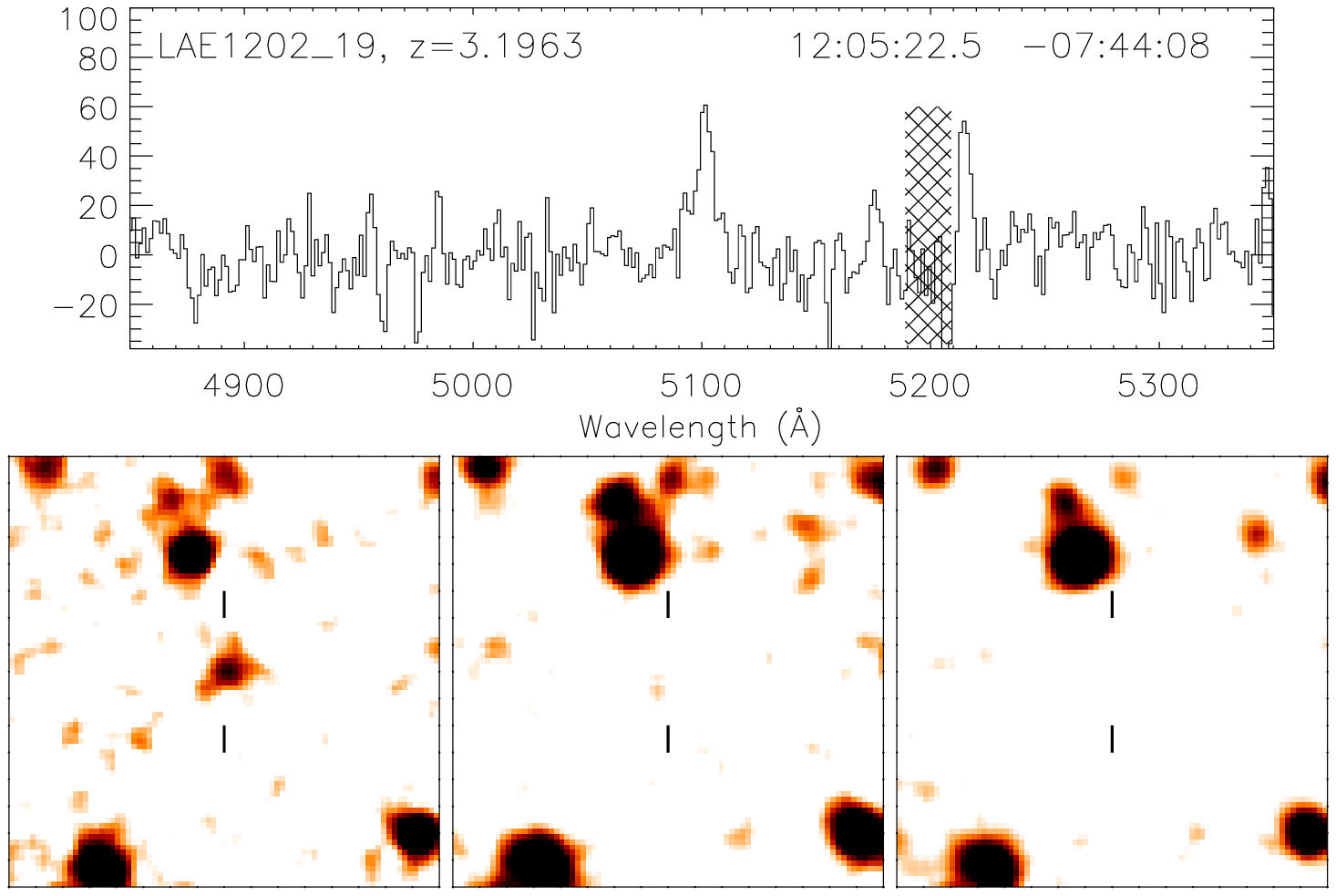}}
\resizebox{0.3\textwidth}{!}{\includegraphics{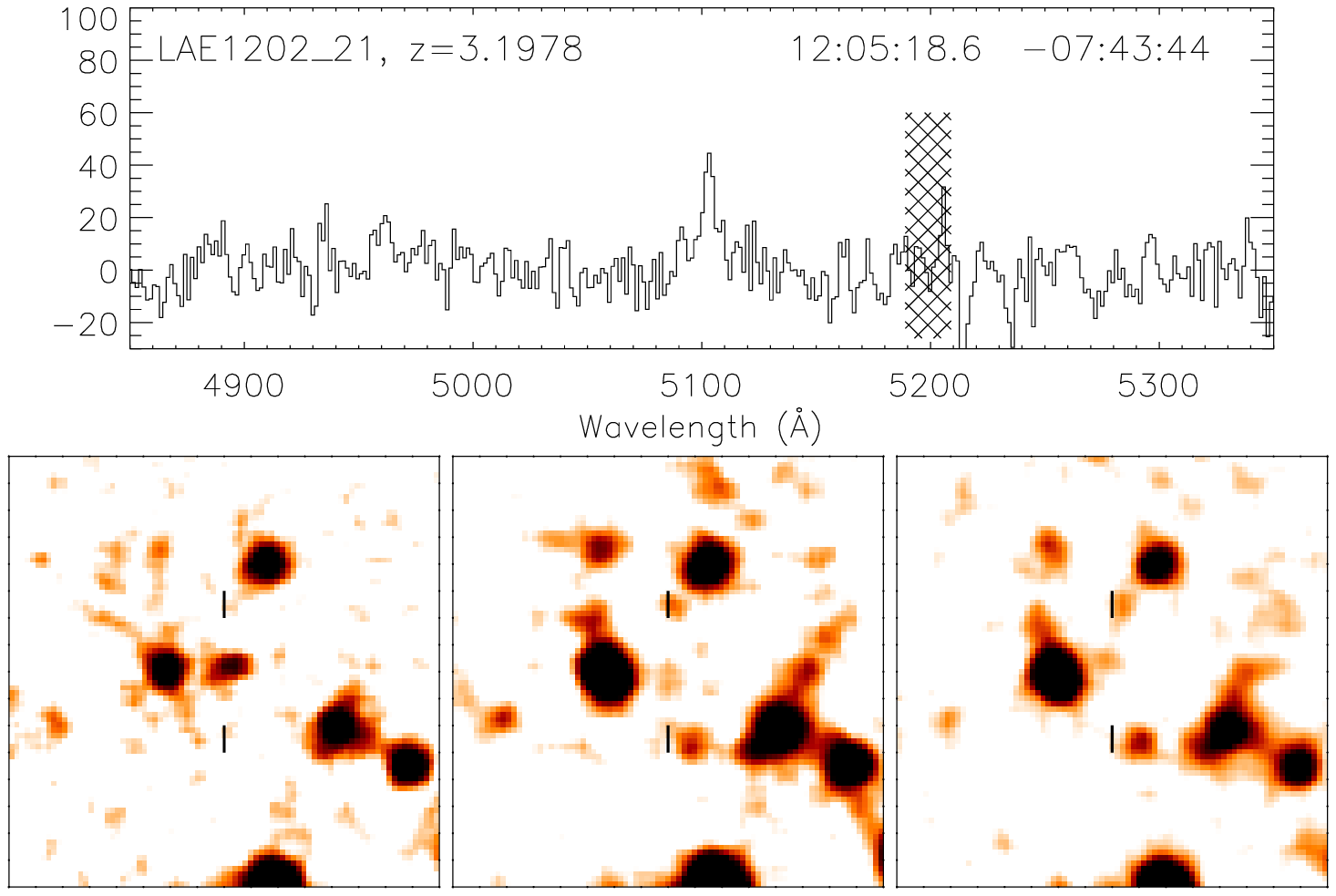}}
\resizebox{0.3\textwidth}{!}{\includegraphics{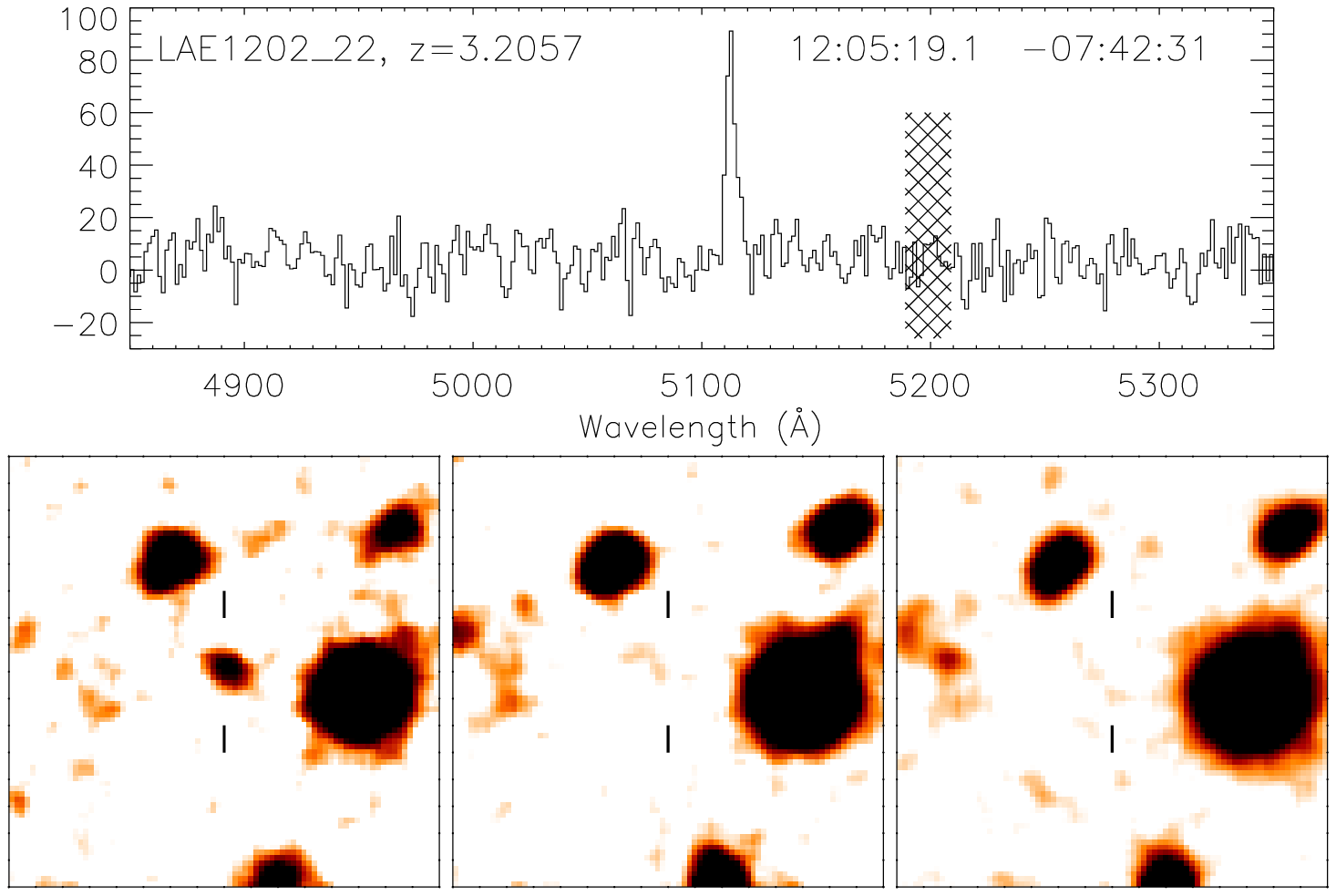}}
\resizebox{0.3\textwidth}{!}{\includegraphics{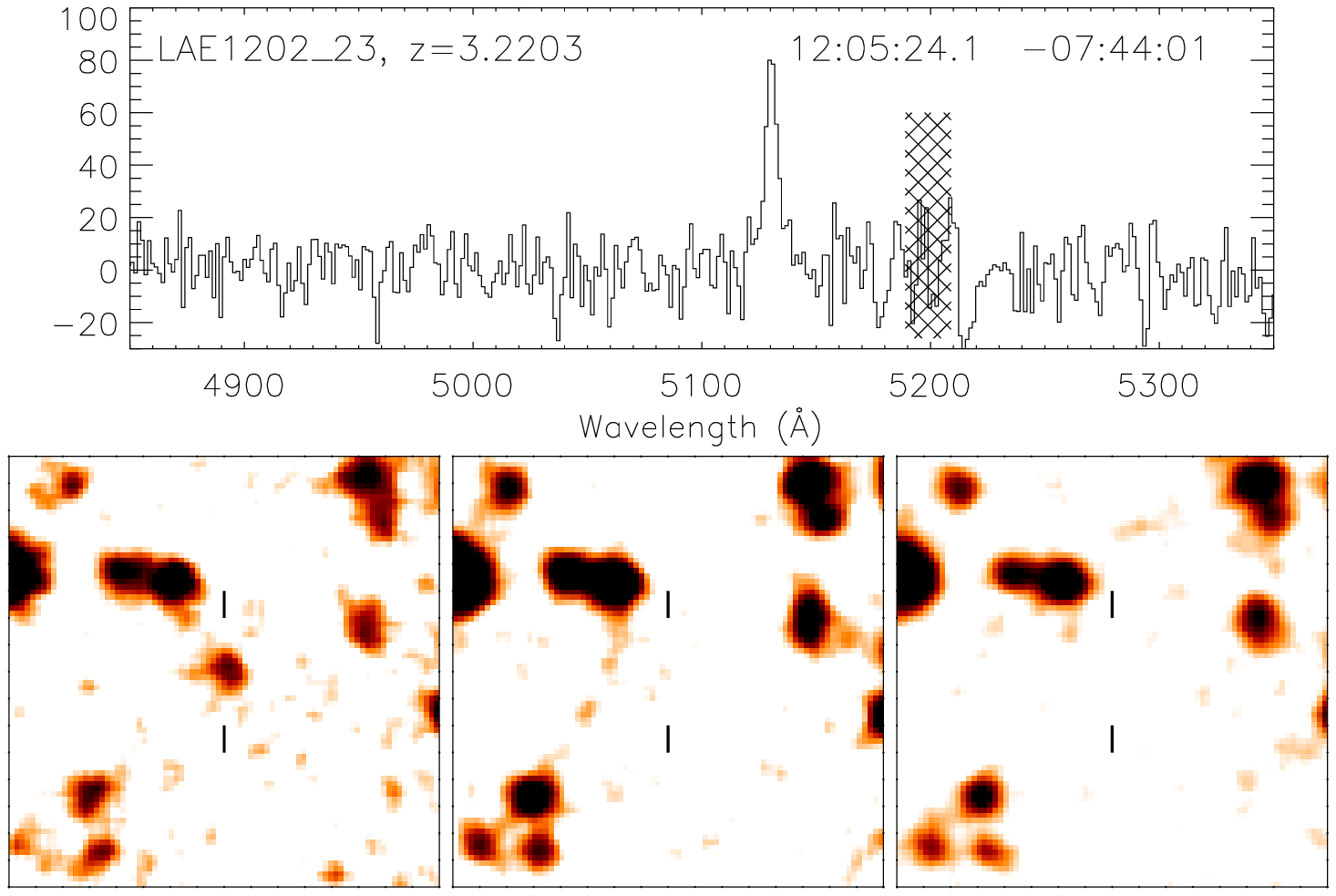}}
\end{center}
\caption{16$\times$16 arcsec$^2$ images and 1-D spectra of 18 (note
that candidates \# 8 and 11 are covered by the same spectrum)
confirmed LAEs in the field of BRI\,1202$-$0725. The hatched areas have
been excluded from the analysis due to the overlap with night-sky
lines. The units on the ordinate of the spectra are counts in 1800
sec. The name, redshift and coordinates (Epoch 2000) are provided for
each object. For each candidate, we show images from the $N$-, $V$-
and $R$-band filters (from left to right) with North up and East to
the left. }
\label{fig:conf_lego}
\end{figure*}

Of the 25 targeted LAE candidates, 18 are confirmed emission line
objects. For the seven remaning candidates no emission line was
identified in the spectrum. The spectra for the 18 confirmed
candidates are shown in Fig.~\ref{fig:conf_lego}. We consider a
candidate confirmed if there is an emission line detected with at
least $3\sigma$ significance at the correct position in the slitlet
within the wavelength range corresponding to the filter
transmission. All of the confirmed emission line candidates are
most likely to be Ly$\alpha$ based on the absence of other emission
lines at longer wavelengths (mainly \ion{NeIII}, \ion{H$\beta$} and
\ion{O[III]}, the positions of which are all covered by our
spectroscopy). The limits on the flux ratios we infer are similar to
those derived in Fynbo et al.\ (2001) and based on their analysis
contamination from OII emitters is very unlikely.  The overall
efficiency for detection and confirmation of LAEs is therefore
\#(confirmed LAEs)/\#(observed LAEs)=18/25=72\%. This is the same as
found for Q\,2138$-$4427 and somewhat less than the results of
BRI\,1346$-$0322 (Paper~I).

\begin{figure}
\resizebox{\columnwidth}{!}{\includegraphics{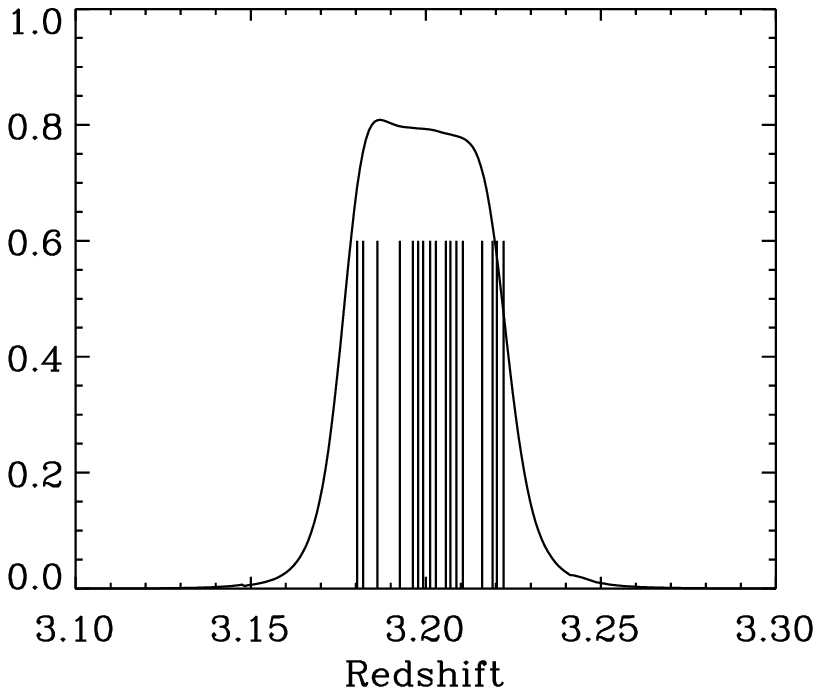}}
\caption{Redshift distribution of LAEs in the field of BRI\,1202$-$0725
relative to the filter transmission curve. The redshifts of LAEs fill
out the volume probed by the filter.}
\label{fig:zdist_LAEs}
\end{figure}

The redshift distribution for the LAE sample in the field of
BRI\,1202$-$0725 is shown in Fig.~\ref{fig:zdist_LAEs} and compared with the
filtercurve. It can be seen that the targets fill out the volume
probed by the filter. The mean redshift of the LAEs is 3.203 with a
standard deviation of 0.013. 

We have investigated the properties of the non-confirmed candidates
and find that they are in the red end of the broad-band colours of the
candidate sample and unresolved. However, their low line fluxes
prevent them from being confirmed by the present spectroscopic
observations. This analysis was carried out for all three fields and
we found similar properties in all cases.

\section{The LAE population}

In the following we combine the results of the entire Building the
Bridge Survey to characterise the population of LAEs. The survey
constitutes observations of three fields centered on the quasars
BRI\,1202--0725, BRI\,1346--0322 and Q\,2138--4427 searching for
Ly$\alpha$ emitters at redshifts $z=3.20, 3.15$ and 2.85,
respectively. These redshifts are targeted due the presence of high
column density absorption systems along the line of sight towards the
quasars. In the fields 25, 26 and 36 emission line objects were
identified through the narrow-band technique. In each of the fields of
BRI\,1346--0322 and Q\,2138--4427 the emission from two emitters were
identified with other emission lines than Ly$\alpha$ \citep[in three
cases corresponding to \ion{[OII]}\ and the fourth case \ion{CIV},
][]{fynbo2003}. Therefore, these four objects have been excluded from
the current analysis.  In the following analysis we consider 18, 18
and 23 spectroscopically confirmed LAEs as the confirmed sample. The
entire photometric sample includes 7, 6 and 11 additional
objects. These have not been confirmed spectroscopically. For the
fields of BRI\,1202--0725 and Q\,2138--4427 all candidates were
observed, so the non-confirmed systems were lacking an emission line
at our sensitivity. For the field BRI\,1346--0322 three candidates
were not observed leaving three as spectroscopically not confirmed
candidates.

\begin{table*}
\caption{Overview of LAE samples from the three fields included in
the present survey.}
\label{tab:refvalues}
\begin{minipage}{\textwidth}
\begin{center}
\begin{tabular}{lrcccccccccc}
\hline\hline
Field & z & $N_{tot}$\footnote{Number of LAEs in the field. In the fields of BRI~1346--0275 and Q~2138--4427 two foreground emission line systems have been excluded from this sample.} &  $N_{conf}$\footnote{Number of confirmed LAEs in the field.} & $N_{non}$\footnote{Number of candidates not spectroscopically confirmed.} & Dens.(conf) & EW$_{0,conf}$ & EW$_{0,non}$ & $L_{conf}$(Ly$\alpha$)\footnote{Lower limit luminosity for those candidates that are detected with a significance of at least $1\sigma$ in the total magnitude.} & $L_{non}$(Ly$\alpha$)$^d$ & SFR$_{conf}$ & SFR$_{non}$\\
 & & & & & $arcmin^{-2} \Delta z^{-1}$ & {\AA} & {\AA} & 10$^{41}$ erg s$^{-1}$ &  10$^{41}$ erg s$^{-1}$ & M$_\odot$yr$^{-1}$ & M$_\odot$ yr$^{-1}$\\
\hline
BRI~1202--0275 & 3.20 & 25 & 18 & 7 & 8 & $\geq27$ & $\geq21$ & $\geq6.06$ & $\geq4.07$ & $\geq0.62$ & $\geq0.42$\\ 
BRI~1346--0322 & 3.15 & 24 & 18 & 6 & 8 & $\geq22$ & $\geq 26$ & $\geq12.21$ & $\geq9.92$ & $\geq1.25$ & $\geq1.02$\\
Q~2138--4427 & 2.85 & 34 & 23 & 11 & 10 & $\geq20$ & $\geq18$ & $\geq3.66$ & $\geq 4.58$ & $\geq 0.38$ & $\geq0.47$\\
\hline
\end{tabular}
\end{center}
\end{minipage}
\end{table*}

The LAE population is characterised in terms of magnitudes, colours
and derived entities like Ly$\alpha$ flux, luminosity, EWs and star
formation rates (SFRs). For total magnitudes we use the SExtractor
MAG\_AUTO which are used to compute the Ly$\alpha$ flux, luminosity
and SFR. The colour indices are computed based on isophotal magnitudes
(MAG\_ISO) and are used to estimate the EW. For all magnitudes the
same detection area is used in all bands. Details of the computation
of the derived properties can be found in \cite{fynbo2002}.  Finally,
in the next section we derive the luminosity function of LAEs at
$z\sim3$ and compare this with the recent measurements by
\cite{gronwall2007,ouchi2008} and
\cite{rauch2008}. Table~\ref{tab:refvalues} gives an overview of the
LAE samples of the three fields. From the table it can be seen that
the survey covers LAEs with luminosities in Ly$\alpha$ down to a few
times $10^{41}$ ergs\; s$^{-1}$, making it one of the deepest surveys of
Ly$\alpha$ emitters at z$\approx3$.  Table~\ref{tab:comp_samples}
summarises the main characteristics of our sample and the three used
for comparison.

\begin{table*}
\caption{Summary of the main properties of our and the comparison samples.}
\label{tab:comp_samples}
\begin{minipage}{\textwidth}
\begin{center}
\begin{tabular}{llrrrr}
\hline \hline
Ref. & Type & z & N & Area & Ly$\alpha$ luminosity \\
 & & & & sq.arcmin & ergs\; s$^{-1}$ \\
\hline
This sample & Spectroscopic/photometric&  $\sim$3.0 & 59/83 & 133 & $\gtrsim$0.3$\cdot$10$^{42}$\\
& Around absorption systems\\
Serendipitous & Spectroscopic & $2.30\leq z\leq 3.61$ & 6 & $-$ & $-$ \\
& Field\\
\cite{gronwall2007} & Photometric only& $\sim$3.1 & 162 & 1008 & $\gtrsim1.3\cdot$10$^{42}$\\
 & Field\\
\cite{ouchi2008}\footnote{These authors have also carried out spectroscopic observations for a subsample of 41 candidates.} & Photometric only & $\sim3.1$ & 356 & 3538 & $\gtrsim10^{42}$\\
& Field\\
\cite{rauch2008} & Long-slit spectroscopy & $ 2.67\leq z\leq 3.75 $ & 27 & $-$ & $(0.015-1.73)\cdot10^{42}$\\
& Field\\
\hline
\end{tabular}
\end{center}
\end{minipage}
\end{table*}

\subsection{Characteristics of spectroscopically confirmed LAEs}
\label{sec:char-LAEs}

\begin{figure}
\begin{center}
\resizebox{\columnwidth}{!}{\includegraphics{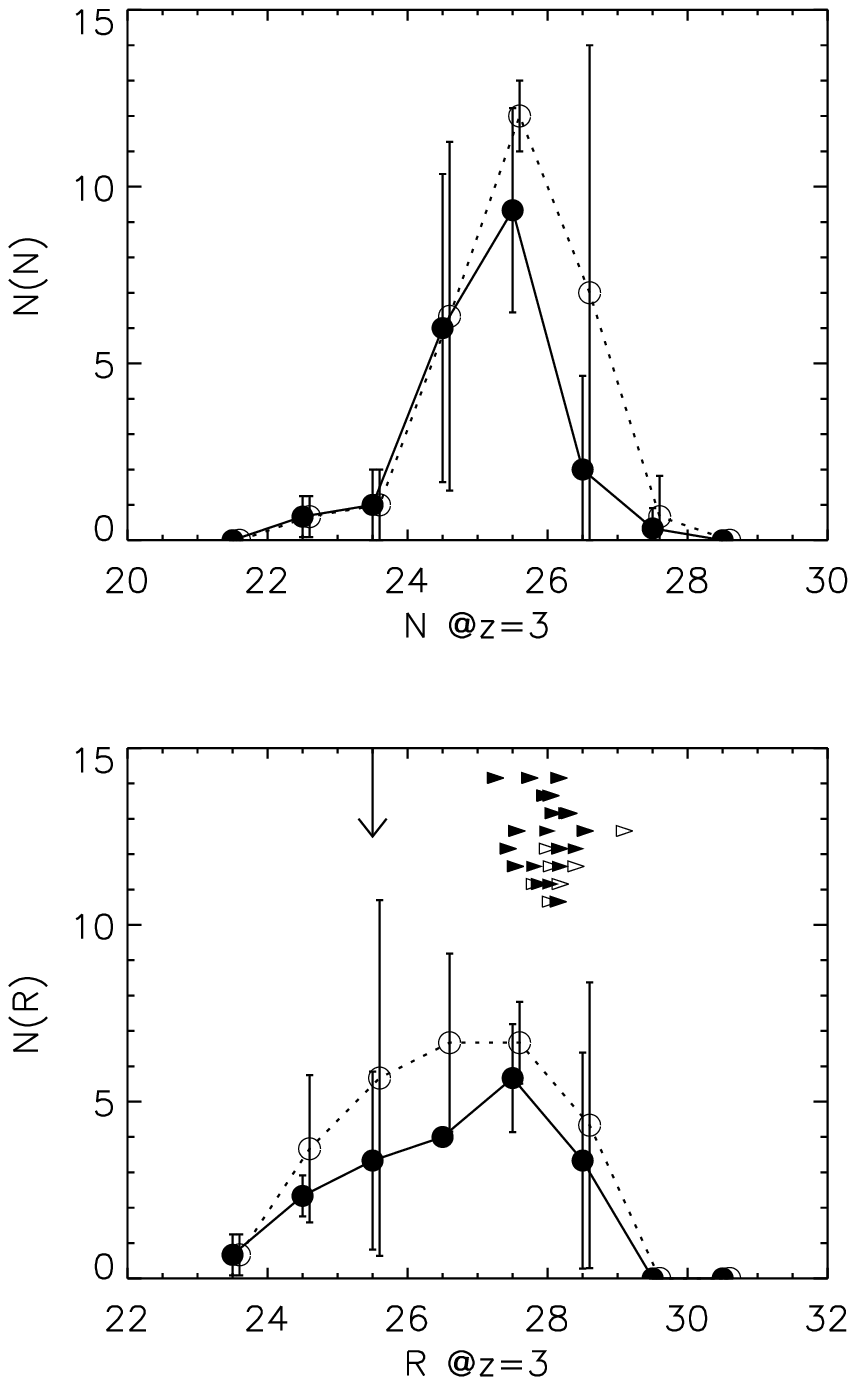}}
\caption{Total magnitude (MAG\_AUTO) distribution of the LAEs in
Narrow- (top) and $R$-band (bottom). The curves are averages between
fields and error bars are the corresponding standard deviations. Open
circles and dotted lines mark the distribution for all photometrically
selected candidates, while filled symbols and solid lines mark the
confirmed Ly$\alpha$ emitters. The triangles indicate $1\sigma$ upper
limits for galaxies detected at lower levels. The arrow in the lower
plot marks the $R=25.5$ limit for spectroscopic surveys of absorption
line systems. It should be noted that for the candidate selection the
isophotal magnitudes were used and here we use total magnitudes. All
magnitudes have been rescaled to $z=3$, though neglecting the small
differences in k-correction for the $R$-band.   }
\label{fig:magdist}
\end{center}
\end{figure}

In
Tables~\ref{tab:lego_properties_1202}--\ref{tab:lego_properties_2138}
we give the measured and derived properties for the confirmed
LAEs. The magnitudes in these tables are total magnitudes, and the
lower limits correspond to $1\sigma$ significance levels for the
SExtractor MAG\_AUTO apertures. These translate into the limits given
for the fluxes and luminosities. The EWs are derived from the colour
indices based on the isophotal magnitudes. In this case the limits are
caused by a less significant detection in the on-band broad ($B$ or
$V$) and the values correspond to using the $1\sigma$ levels for these
bands.

From Table~\ref{tab:refvalues} it can immediately be seen that the
total number of objects is comparable among the three fields. The
detected number of LAEs translates into surface densities of 8, 8 and
10 per square arcmin per unit redshift, respectively for the fields of
BRI~1202--0275, BRI~1346--0322 and Q~2138--4427, consistent with the
almost similar flux limits in the three fields.  From the table it can
also be seen that the other properties are similar between the fields
and between the confirmed LAEs and non-confirmed candidates.

In Fig.~\ref{fig:magdist} we show the magnitude distribution for the
LAEs. The lines correspond to candidates with measured magnitudes
while the triangles indicate systems with $1\sigma$ upper limits. The
observed magnitudes are translated to correspond to a redshift $z=3$
by correcting by the difference in the distance modulus between the
observed redshift and $z=3$. For the $R$-band we do not apply any
K-correction since it is a negligible effect over such small redshift
changes. The errorbars are standard deviations among the fields and
indicate that even though the results in the three fields are
consistent, small numbers and cosmic variance leads to significant
differences between the fields. For the narrow-band it is clear that
all the bright objects have been confirmed to correspond to Ly$\alpha$
emitters while the non-confirmed cases are in general found in the
faint end of the distribution.

From the magnitude distribution of the $R$-band it can be seen that
all the LAE candidates have very faint magnitudes. For absorption line
systems the magnitude limit of achieving a redshift is $R$$\sim$25.5
indicated by the arrow in the figure. Therefore, we note that of the
confirmed emitters only 6 out of 59 or $\lesssim$10\% are brighter
than this limit. The fraction is similar for the complete candidate
sample. This is consistent with the non-confirmed candidates being
spread over all $R$-band magnitudes. The faint $R$-band magnitudes of
the galaxies emphasises their importance for tracing the faint end of
the luminosity function, possibly contributing a significant fraction
of the total star formation \citep[see also ][]{reddy2008}.

\begin{figure}
\begin{center}
\resizebox{\columnwidth}{!}{\includegraphics{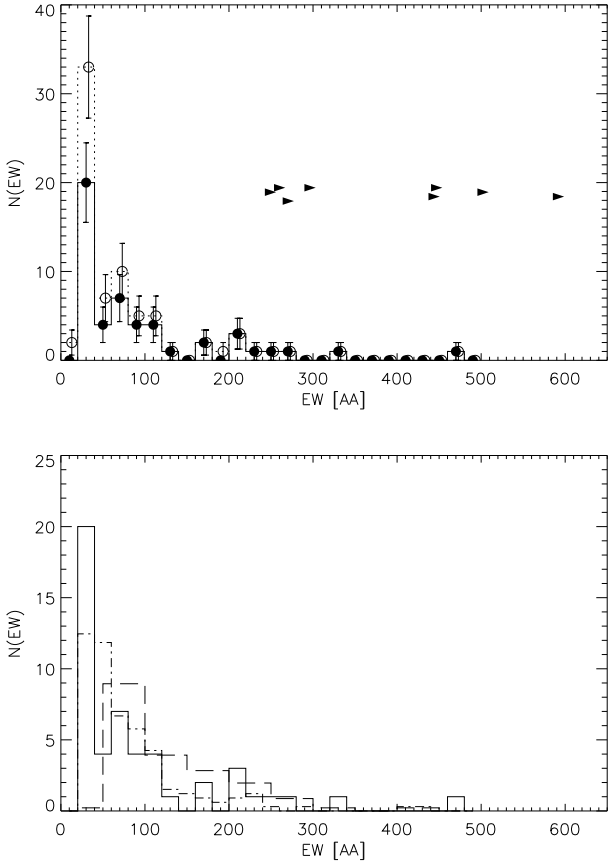}}
\caption{Rest frame equivalent width distribution for the confirmed
(solid symbols and line) and entire (open symbols and dashed line)
samples (upper panel). In the lower panel the distribution for the
confirmed sample is compared with that of \cite[][dot-dashed
line]{gronwall2007} and \cite[][dashed line]{ouchi2008}. The triangles
indicate cases for which only lower limits (at the $1\sigma$ level)
could be determined.}
\label{fig:ew}
\end{center}
\end{figure}

Fig.~\ref{fig:ew} shows the distribution of rest frame equivalent
widths derived from the colour index ``narrow band minus on-band
broad'' isophotal magnitudes. We have not corrected the broad-band
magnitudes for the narrow-band contribution. Our measurements are
compared with those of \cite{gronwall2007} and \cite{ouchi2008}. It
can be seen that the sample of confirmed LAEs for which we could
measure the EW reliably is consistent with our survey being deeper
than the comparison samples. It can also be seen that the Ouchi et
al. sample has a higher fraction of emitters with high EW than the
Gronwall et al sample. Our sample it is more consistent with the Ouchi
et al. than with the Gronwall et al. sample.

Table~\ref{tab:refvalues} also lists the range of SFR derived from the
Ly$\alpha$ luminosities (following Fynbo et al.\ 2002). The SFRs are 
found to be in the range
$0.38M_\odot$yr$^{-1}$ to $31M_\odot$yr$^{-1}$ assuming negligible
extinction.

\section{Luminosity function of LAEs at $z\sim3$}

A widely used diagnostic for describing galaxy samples is their
luminosity function (LF). Here, we derive the LF of LAEs at $z=3.0$
combining the data from the present work and from Paper~I. The
complete survey encompasses three fields with a total of 59
spectroscopically confirmed LAEs. For each field we derive the LF
independently to take into account the different narrow-band filters
and incompleteness functions. Finally, the individual results are
combined to provide the LF at $z=3.0$.

%A major challenge is to estimate the selection function of the LAE
%catalogues described in the previous sections and in Paper~I. Due to
%the variations in the observations we estimate the incompleteness for
%each field and thus derive the LF independently. Finally, we combine
%all three LFs to improve the statistics of the final function.

\subsection{Estimating the incompleteness correction}

\begin{figure}
\begin{center}
\resizebox{\columnwidth}{!}{\includegraphics{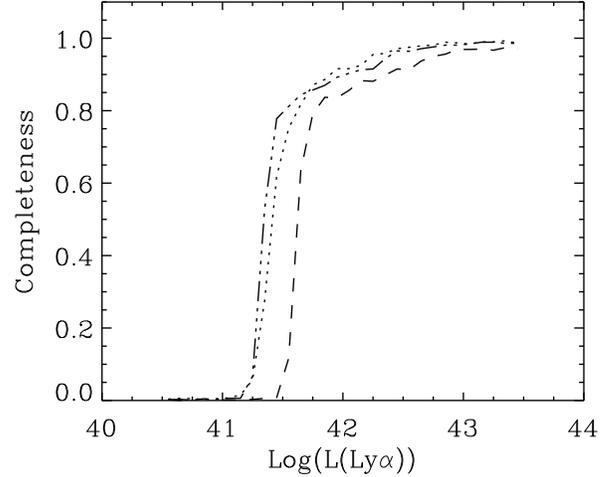}}
\caption{Completeness levels as function of Ly$\alpha$ luminosity. The
dotted line corresponds to the field of BRI~1202--0725; dashed line to
BRI~1346--0322 and triple dot-dashed line to Q~2138--4427.}
\label{fig:phot_selfct}
\end{center}
\end{figure}

When constructing the sample of LAEs the main source of incompleteness
comes from the photometric selection of candidates. The following
process is carried out separately for each field to take into account
the specific properties of the different data sets. To estimate the
detection completeness of the photometric samples we distribute 100
artificial objects in the images and assign broad- and narrow-band
magnitudes in a range covering the observed values as well as an
additional margin. The magnitudes were assigned based on the
off-broad band. In this band magnitudes are distributed uniformly in
the interval 5 to 35. For the narrow- and on-broad bands magnitudes
were assigned based on a uniform colour index (``narrow $-$
off-broad'' and ``on-broad $-$ off-broad'') distribution in the
interval minus five to five. This results in uniform magnitude
distributions in the narrow- and on-broad bands in the magnitude
interval 10 to 30 dropping off and vanishing at magnitudes of zero and
40.  The objects are modeled with a two-dimensional gaussian shape
with the size corresponding to the seeing measured in the individual
images. Each artificial object is added to each image scaling the
gaussian to the appropriate luminosity depending on the magnitudes
computed individually for each band. This process ensures good
coverage for both magnitudes and colour indices. The procedure is
repeated 1000 times for each field, thus a total of 10$^5$ artificial
objects are used in the estimate. The images with artificial objects
are now treated exactly as the original images and the recovery rate
yields the detection completeness. The completeness functions
estimated for the three fields in the narrow-band filters are shown in
Fig.~\ref{fig:phot_selfct}. From that figure it can be seen that the
data of the field of BRI~1346--0322 are slightly shallower than the
other two fields, which are very similar. These completeness functions
are used to correct the measured luminosity functions below. Note that
we have not corrected our data for the effects of a non-square
band-pass or the photometric error function \citep[see
][]{gronwall2007}.

A possible concern is whether the extension of the Ly$\alpha$
emitters have a significant effect on the estimated completeness
functions. To test this we have carried out a complementing test where
the original background subtracted images were first dimmed. Then we
corrected the background noise to the original level and tried to
recover the already detected candidate emitters. We dimmed the images
by 0.25, 0.5 and 0.75 magnitudes. The completeness functions estimated
in this way were similar to the ones described above, but suffer from
poor statistics (only 80 sources in total were found in the original
images) and only tracing the magnitudes and colour indices of the
detected sources. This makes the previously described functions more
reliable for correction purposes and are the ones used in the
following.

\subsection{Ly$\alpha$ luminosity function}

\begin{figure}
\begin{center}
\resizebox{\columnwidth}{!}{\includegraphics{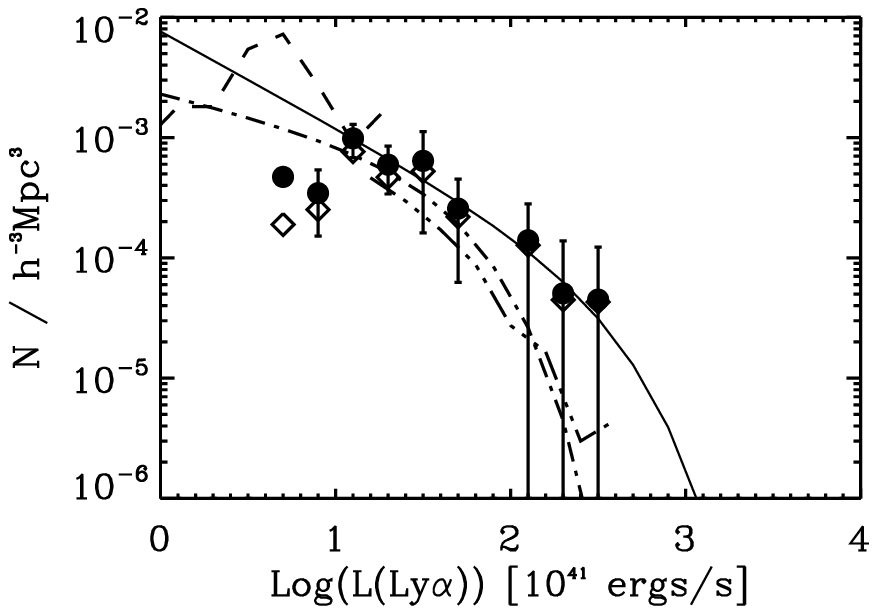}}
\resizebox{\columnwidth}{!}{\includegraphics{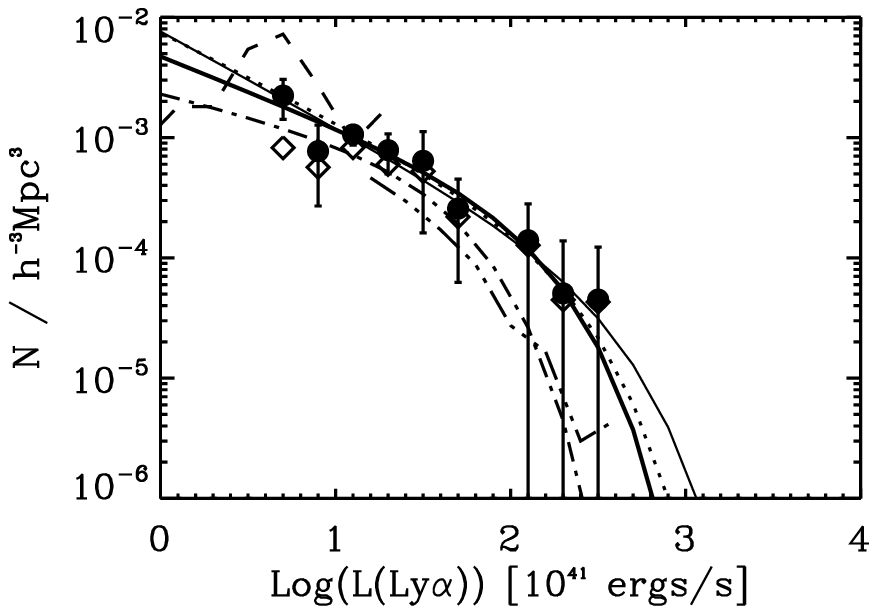}}
\caption{The derived differential Ly$\alpha$ luminosity function for
the sample of confirmed (upper) and all (lower) LAEs. The points with
open symbols are not corrected for incompleteness, while those with
solid symbols are. The thin solid line indicate our fit to the
confirmed candidates. The thick solid and dotted lines in the lower
panel mark the fits to the deep and brighter sample of all cadidates,
respectively. The error bars are obtained from the
$1\sigma$-error bars of the individual fields by standard error
propagation.} The fits are described in detail in the text. Our result
is compared with the luminosity function for emitters at $z=3.1$ by
\cite{ouchi2008} in triple-dot dashed line, \cite{gronwall2007} in
dot-dashed line and \cite{rauch2008} in dashed line.
\label{fig:lya_lf}
\end{center}
\end{figure}

We derive the differential luminosity function for the confirmed LAEs
with a simple classical approach that has been used in many previous
works
\cite[e.g. ][]{ouchi2003,ajiki2003,hu2004,malhotra2004,ouchi2008}. The
volume used to convert the observed number of systems to volume
densities is computed based on the field size and having a depth
corresponding to the FWHM of the narrow-band filter. Furthermore, we
correct the observed luminosity function by the completeness function
estimated in the previous section to account for the incompleteness of
the sample due to our target selection. The obtained luminosity
function is shown in Fig.~\ref{fig:lya_lf}. The error-bars are
derived from the errors in the individual fields by standard error
propagation.

We fit our derived LF with the Schechter function \citep{schechter76}
using a $\chi^2$-minimisation. We carry out three fits in
total. First, we fit the sample of confirmed candidates with $L^*
>=10^{42}\mathrm{ergs\; s^{-1}}$. Second, we fit the entire photometric
sample in two luminosity intervals: $L^* >=10^{42}\mathrm{ergs\; s^{-1}}$ and
$L^* >=10^{41.6}\mathrm{ergs\; s^{-1}}$. The brighter limit corresponds to
the region where all fields are estimated to be close to complete,
while the second limit is where the shallowest field is essentially
not contributing anything (see Fig.~\ref{fig:phot_selfct}). The
results of the fits are summarised in Table~\ref{tab:fit_res}.  In
Fig.~\ref{fig:lya_lf} we include the fit to the confirmed candidates
as the thin solid line in both panels and the fit to the deep sample
of all candidates is included in the lower panel as the thick solid
line. The fit to all candidates with the brighter luminosity limit is
included as the dotted line. It can be seen that despite the small
differences in the best fit parameter values the functions are
consistent.

\begin{table*}
\caption{Results of fitting a Schechter function to the LF.}
\label{tab:fit_res}
\begin{center}
\begin{tabular}{lllll}
\hline\hline
Type & $\log(L^*)$ & $\alpha$ & $\phi$ & $\chi^2$\\
& (ergs\; s$^{-1}$) & & $h^{-3}Mpc^3$ & \\
\hline

\vspace{2mm}

Confirmed, $\log(L^*)>=42$ & $43.55_{-0.10}^{+0.05}$ & $-1.80_{-0.06}^{+0.08}$ & $1.5_{-0.3}^{+0.3}\cdot 10^{-4}$ & 0.32\\

\vspace{2mm}

All, $\log(L^*)>=42$ & $43.30_{-0.05}^{+0.05}$ & $-1.74_{-0.04}^{+0.06}$ & $3.2_{-0.5}^{+0.5}\cdot 10^{-4}$ & 0.28\\

\vspace{2mm}

All, $\log(L^*)>=41.6$ & $43.15_{-0.05}^{+0.05}$ & $-1.58_{-0.04}^{+0.06}$ & $5.8_{-0.8}^{+0.8}\cdot 10^{-4}$ & 2.06\\

\hline
\end{tabular}
\end{center}
\end{table*}

We compare our result with those of \cite{gronwall2007,ouchi2008} and
\cite{rauch2008}. At the faint end we find a good agreement, in
particular if all the non-confirmed candidates are LAEs. At the bright
end we find a marginal excess of objects compared to the other
surveys, even though the numbers are consistent within the error
bars. The (marginal) excess of bright objects may be caused by enhanced 
clustering of 
bright LBGs around QSO absorber fields \citep{bouche2004,bouche2005}.
Comparing the density of galaxies brighter than the
limit of \cite{gronwall2007} of $13\cdot 10^{41}\rm{ergs\; s^{-1}}$ we find only
a marginal excess of a factor of $\sim1.25$. %Such an average
%overdensity is likely the result of targeting the vicinity of known
%structures (DLAs). 
It is also interesting to compare these
environments with the results of \cite{venemans2007} studying radio
galaxies. The radio galaxies traces the largest overdensities, likely
representing proto-clusters. The overdensity of LAEs around radio
galaxies are found to be of the order 2-5. Thus the overdensities
found in our fields are, as expected, smaller. In the field of
Q~2138--4427 the redshifts are concentrated relative to the width of
the filter function (Paper~I, see also \cite{monaco2005}). In this 
field the overdensity is only a
factor of 1.1 with respect to the results of \cite{gronwall2007}, thus
we do not consider this good evidence for a proto-cluster.

\section{Galaxy counterparts of the QSO absorbers}

\begin{figure}
\centerline{\hbox{
\psfig{figure=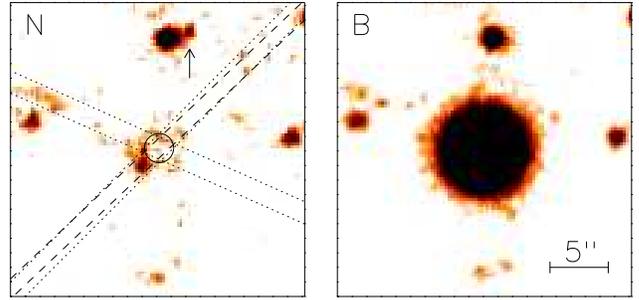,width=8.4cm,clip=,bbllx=6.pt,bblly=2.pt,bburx=446.pt,bbury=214.pt,angle=0.}}}
\caption[]{FORS\,1 narrow-band image centered on Q\,2138$-$4427 (left
panel). North is up and East is to the left. The wavelength at peak
transmission of the (He{\sc ii}) narrow-band filter used corresponds
to Ly$\alpha$ emission at $z=2.85$. The PSF of the QSO was
carefully subtracted which reveals a source only $1\farcs 4$ away from
the QSO line-of-sight. The different orientations of the slits during
the MOS observations (Paper~I) and the complementary deep long-slit
spectroscopic follow-up are shown by, respectively, dotted and dashed
lines. The arrow in the upper part of the image marks the location of
the closest LAE (LEGO2138\_36) that was selected by
comparing the narrow-band image with the FORS\,1 $B$ and $R$
broad-band images. In the right panel, the $B-$band image is shown
with the same scale and orientation as the narrow-band image.}
\label{candidate}
\end{figure}

One of the goals of our survey was to bridge the gap between emission
and absorption selected objects. The three studied fields are centered
on QSOs with high column density absorption systems at the redshift
for which Ly$\alpha$ falls in the narrow-band filters. In the
fields of BRI\,1346$-$0322 and BRI\,1202$-$0725 the absorbers are
Lyman-limit systems and in the field of Q\,2138$-$4427 it is a DLA.
The observed flux of Q\,2138$-$4427 is reduced by approximately 1.5
magnitude in the narrow-band image as shown in Fig.~3 of Paper~I. This
is due to the strong DLA line that absorbs the QSO light making it
much easier to search for any faint emission close to the QSO in the
narrow band than in the broad-band images. Moreover, if the galaxy
counterpart of the $z_{\rm abs}=2.852$ DLA system has Ly$\alpha$
emission, it would be relatively brighter in the narrow-band than in
the broad-band images. To subtract the image of the QSO in the
narrow-band image, we used PSF-subtraction. We modelled the PSF of the
QSO from ten bright, yet unsaturated, stars present in the field. The
extension ALLSTAR of the DAOPHOT-II software
\citep{stetson1987,stetson1994} was used to perform the final PSF
model fitting and subtraction.  After careful PSF subtraction, we
detected an extended source at an impact parameter of $1\farcs 4$
(corresponding to 12 kpc at $z=2.85$) from the QSO line-of-sight at a
position angle of $136.5^o$ (see Fig.~\ref{candidate}). The proximity
of this source to the QSO line-of-sight makes it a good candidate for
the DLA galaxy counterpart.

\begin{figure}
\centerline{\hbox{
\psfig{figure=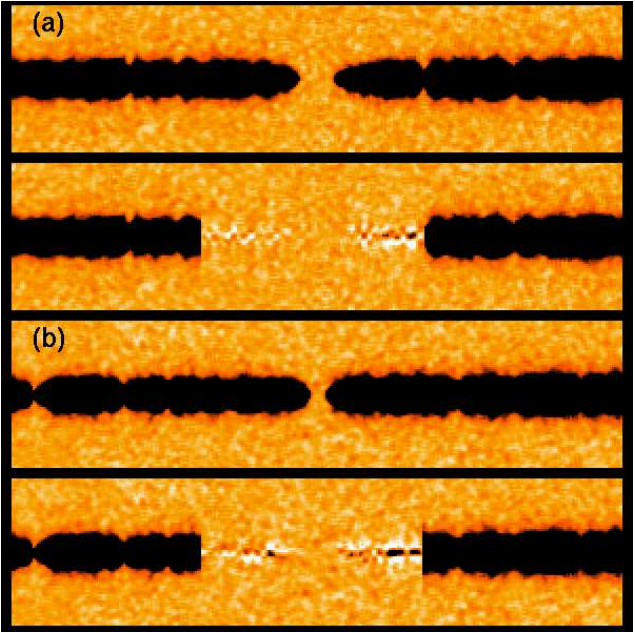,width=8.4cm,angle=0.}}}
\caption[]{(a) portion of the 2-D long-slit spectrum of Q\,2138$-$4427
centered on the Ly$\alpha$ trough of the DLA system at $z_{\rm
abs}=2.852$ (first top panel). In the second top panel, the PSF of the
QSO has been subtracted using the Spectral PSF II optimal extraction
algorithm \citep{moeller2000}. (b) same as above for the DLA system at
$z_{\rm abs}=2.383$ toward Q\,2138$-$4427 (bottom two panels). The
host galaxies of the DLA systems are detected in neither case. The
object seen in the narrow-band image shown in Fig.~\ref{candidate}
must be a continuum source.}
\label{nondetect}
\end{figure}

To establish whether the source seen in Fig.~\ref{candidate} is the
galaxy counterpart of the $z_{\rm abs}=2.852$ DLA system, we gathered
spectra of Q\,2138$-$4427 using both FORS\,1 Multi-Object Spectroscopy
(MOS) and long-slit spectroscopy. The details of the MOS observations
were given in Paper~I. The positions of the two MOS slitlets that
covered the QSO are marked with dotted lines in the left panel of
Fig.~\ref{candidate}. One of the $1\farcs 4$-wide MOS slitlets covered
both the QSO and the candidate DLA galaxy counterpart while the second
MOS slitlet was oriented at a position angle about $70\degr$
smaller. None of these spectra showed Ly$\alpha$ emission from the
DLA absorber.  However, the atmospheric conditions during the MOS
observations were quite poor (see Paper~I) and we consequently
obtained a deeper ($\sim 5$ hr total integration time) FORS\,1
long-slit spectrum with a $1\farcs 0$-wide slit and the G600\,B grism
covering the candidate under very good seeing conditions (${\rm
FWHM}=0\farcs 71$ in the combined spectrum). The position of the long
slit is shown with dashed lines in Fig.~\ref{candidate}. The 2-D
long-slit spectrum is displayed in the upper two panels of
Fig.~\ref{nondetect}. We used the spectral PSF II optimal extraction
algorithm \citep{moeller2000} to subtract the emission from the QSO in
the wings of the DLA trough. No Ly$\alpha$ emission from the
candidate DLA galaxy counterpart is detected.

We conclude that the source seen in Fig.~\ref{candidate} is not an LAE
at $z=2.85$. Possible Ly$\alpha$ emission from this source should be
fainter than the $3\sigma$ very low detection limit of $\sim 4\times
10^{-18}$ erg s$^{-1}$ cm$^{-2}$. Within $1\arcsec$ from the QSO
line-of-sight, our detection limit for Ly$\alpha$ emission is even
lower (by less than a factor of two though) due to the combined limits
from both the MOS and long-slit spectra. The nearest confirmed LAE is
LEGO2138\_36 (Paper~I) and it is situated at a distance of $8\arcsec$
from the QSO line-of-sight. This means that if the $z_{\rm abs}=2.852$
DLA galaxy counterpart is an LAE it should actually be fainter than
all of the LAEs detected in our survey.

What could be the origin of the source seen in Fig.~\ref{candidate}?
One possibility is that this source is associated to the galaxy
counterpart of the lower redshift, $z_{\rm abs}=2.383$ DLA system
toward Q\,2138$-$4427. In this case the emission detected in the
narrow-band image would be continuum emission.  Making this
assumption, we estimate a broad-band magnitude of
$B$(AB)$\approx N$(AB$)=25.7$ for this object.  In the lower two panels
of Fig.~\ref{nondetect}, we show the 2-D spectrum of the QSO around
the corresponding DLA trough.  No Ly$\alpha$ emission is detected
at more than the $5\sigma$ confidence level. There are two $3\sigma$
peaks in the spectrum however but we consider these too uncertain to
allow further discussion.  Another interesting possibility is that the
source is part of the host galaxy of Q\,2138$-$4427.

In none of the other two fields do we see any evidence for a narrow
band source close to the QSO position after PSF subtraction. Note,
however, that for these sight lines the intervening absorbers are only
Lyman-limit systems so the fluxes from the QSOs are much less reduced
by the absorption systems than for Q\,2138$-$4427.

\subsection{Serendipitously detected LAEs}

\begin{figure}
\centerline{\hbox{
\psfig{figure=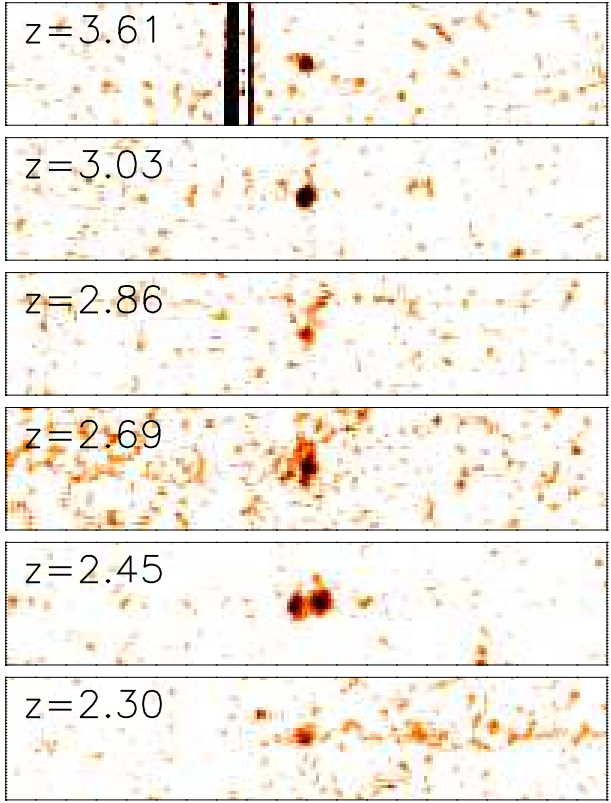,width=8.4cm,angle=0.}}}
\caption[]{2-D spectra of probable LAEs serendipitously observed in the
FORS\,1 long slit spectrum.
The redshifts range from $z=2.30$ to 3.61. The $z=2.86$ object is LAE2138\_12
previously observed also in one of the MOS slitlets (see paper~I).}
\label{serendip}
\end{figure}

The combined FORS\,1 2-D long-slit spectrum covers a solid angle of
$370\times 1$ arcsec$^2$ and the wavelength range from 3600 \AA\ to
6000 \AA . This spectrum is therefore sensitive to LAEs with redshifts
in the range $2.0<z<3.9$. Assuming a constant source density of
approximately ten LAEs per arcmin$^2$ per unit redshift, as is found
in our survey for $z\sim 3$ LAEs down to a flux detection limit of
$N({\rm AB})=26$ (corresponding to a Ly$\alpha$ flux of about
$7\times 10^{-18}$ erg s$^{-1}$ cm$^{-2}$), we expect to
serendipitously observe about six LAEs in the long slit.  This number
is of course only a rough estimate as the sensitivity of the spectrum
is a function of wavelength and the volume density of LAEs down to the
given flux detection limit is likely to be a strong function of
redshift.  In the 2-D spectrum, we actually detect six emission-line
objects with no or very faint continua. These are likely to be LAEs at
redshifts 2.30, 2.45, 2.69, 2.86 (one of which is LEGO2138\_12,
Paper~I), 3.03 and 3.61 (see Fig.~\ref{serendip}). In
Table~\ref{tab:serendip} we give the  redshift and broad-band
magnitudes of the serendipitously detected LAEs. Two of the systems
were not detected in our broad-band imaging. For the others, it can be
seen that their broad-band magnitudes match the typical magnitudes of
the photometrically selected LAEs from our survey. With respect to the
colour indices we find values around zero which is towards the blue
end of the typical candidates.

\begin{table}
\caption{Properties of the serendipitously detected LAEs.}
\label{tab:serendip}
\begin{minipage}{\columnwidth}
\begin{center}
\begin{tabular}{lrrr}
\hline\hline
Id &  $z$ & $B$ & $R$\\
\hline
1 &  2.30 & 26.7 & 26.5\\
2 &  2.45 & 26.2 & 26.1 \\
3 &  2.69 & 27.0 & 27.1\\
4\footnote{This object was detected as LEGO2138\_12 of Paper~I.}& 2.86 & 26.8 & 27.0\\
5\footnote{These objects were not detected in the images and thus no information can be extracted.}& 3.03 \\
6$^b$ & 3.61\\
\hline
\end{tabular}
\end{center}
\end{minipage}
\end{table}

\section{Discussion and Conclusions}

The aim of our survey was to try to bridge the gap between emission
selected and absorption selected galaxies at $z\approx3$. To reach
this goal we have performed the currently deepest narrow-band survey
for Ly$\alpha$ emitting galaxies at $z\approx3$ using narrow-band
imaging at the VLT. Our survey was succesful in establishing the
existence of a large number of galaxies below the flux limit of the
Lyman-break surveys ($R=25.5$). We reach a surface density of LAEs of
the order of 10 per arcmin$^{-2}$ per unit redshift. This is about an
order of magnitude larger than for $R<25.5$ LBGs. It is also about a
factor of two higher than that found by other deep surveys for LAEs at
$z\approx3$ \citep[e.g.][]{gronwall2007}. This difference mainly
reflects the varitation in depth between the surveys, since if we
impose a luminosity limit consistent with that of \cite{gronwall2007}
then we include only 42 of our 83 LAEs consistent which results in
only about 20\% more galaxies in our sample with respect to the
\cite{gronwall2007} results. This marginal overdensity we attribute to
our survey targeting the environments of DLAs and Lyman-limit systems
and not a blank field.

Our survey targeted three fields in which we photometrically selected
89 emission line candidates.  Of these, 63 were confirmed to be
emission line galaxies, however, four of the emission galaxies were
foreground systems. In total we thus identified and spectroscopically
confirmed 59 LAEs in the three fields. Three candidates were not
observed as we could not fit them into the available masks. This
corresponds to a spectroscopic confirmation rate of about 73\% or 69\%
excluding the interlopers. Comparison of the properties of the
spectroscopically confirmed and not confirmed candidates showed that
the non-confirmed cases were mostly in the faint end of the narrow
band magnitude range and any other related quantity (EW, Ly$\alpha$
flux and luminosity) while the broad band properties were similar for
the two groups. Some of the confirmed candidates are only detected at
low signal-to-noise in the spectroscopy, most noteably
LAE1202\_09. This LAE seems to be extended both spatially and in
velocity causing the signal-to-noise in the spectroscopy to be smaller
than for more compact candidates with only marginally resolved lines.
Hence, some of the unconfirmed candidates maybe more broadlined
systems.  We conclude that the non-confirmed systems must be a mix of
spurious candidates and objects with too faint emission lines to be
detected by our spectroscopic follow-up. The properties of the
confirmed sample of LAEs are comparable to that found by other authors
as shown in Figs.~\ref{fig:ew} and \ref{fig:lya_lf}, except being
deeper.  The depth of the $R$-band data does not allow for a detailed
discussion, but about 90\% of the selected emission line galaxies are
fainter than the $R=25.5$ limit for Lyman-break surveys.

Since our survey was started in 2000 we have learned a lot more about
the faint end of the luminosity function at $z\approx3$. The study of
LAEs has progressed substantially both in sample sizes and in the
range of redshifts that have been probed from $z=2$
\citep{fynbo2002,nilsson2008} to $z\approx7$
\citep{iye2006,ota2008}. Also, using Gamma-Ray Bursts (GRBs) it has
been found that a significant fraction of massive stars at these
redshifts die in extremely faint galaxies, e.g. $R\approx$28 for the
host galaxy of GRB030323 at $z=3.37$ \citep{vreeswijk2004} and
$R>29.5$ for the host galaxy of GRB020124 at $z=3.20$
\citep{berger2002,hjorth2003}. A statistical analysis of the
luminosities of GRB host galaxies again points to a large fraction of
the star formation being located at the faint end of the luminosity
function \citep{jakobsson2005,fynbo2008}. Also for continuum selected
galaxies there has been significant progress. \cite{sawicki2006} and
\cite{reddy2008} used the Lyman-break technique to push to
significantly fainter limits confirming an extremely steep faint end
slope.

As for the other (faint) end of the bridge, the study of the galaxy
counterparts of DLAs, there has been disappointingly little
progress. We still only have a few spectroscopically confirmed
counterparts of high-$z$ DLAs \citep{moeller2002,moeller2004,christensen2007}.  
An
interesting development on that issue is the tentative evidence for a
luminosity-metallicity relation at place at $z\approx3$
\citep{moeller2004,ledoux2006} \citep[see
also][]{fynbo2008,pontzen2008}. This implies that targeted searches
for the galaxy counterparts of metal rich DLAs could have
substantially higher success rates than for randomly selected DLAs,
but this remains to be confirmed. It
is also consistent with the nondetection of the galaxy counterpart of
the DLA towards Q2138-4427 as this DLA has a relatively low
metallicity of about [Zn/H]$=-1.74$ \citep{ledoux2006}. Another very
interesting recent discovery is that of extemely faint, extended LAEs
detected spectroscopically by \cite{rauch2008} and argued by the same
authors to be a population of galaxies responsible for the bulk of the
DLAs \citep[see also][]{barnes2008}. The argument appears very
convincing, and if confirmed by more actual detections of DLA galaxy 
counterparts this means that we now have bridged the gap between 
absorption and emission selected galaxies at $z\approx$3. 

Indepedent of the issue of bridging the gap between absorption and emission
selected galaxies it is clear that there is a very numerous population of
high-$z$ galaxies occupying the faint end of the luminosity function. 
There is growing evidence that this population of faint galaxies plays an
important role for many important processes in the early Universe. These
galaxies by far dominate the emission of ultraviolet light
\citep[e.g.,][]{jakobsson2005,fynbo2008} and most likely also the
emission of ionizing radiation \citep{bianchi2001,faucher2008,loeb2008}.
They also most likely contain a large fraction of the total metal budget 
in galaxies and are responsible for a large fraction of the
enrichment of the intergalactic medium at $z\approx3$ 
\citep[e.g.,][]{sommer-larsen2008}. Currently it is extremely difficult
to infer more detailed astrophysical properties (e.g., metallicities,
dust content, stellar populations, masses, etc.) for this class of objects.
In a few cases like DLAs and GRB host galaxies we can infer several of these
properties, but in general we cannot. With the advent of 30m class telescopes
the future looks more promising.

\begin{acknowledgements}
We thank the anonymous referee for useful comments.  We are grateful
to Dr. M. Ouchi and to M. Rauch for providing us with comparison
data. The Dark Cosmology Centre is funded by the Danish National
Research Foundation.  LFG acknowledges financial support from the
Danish Natural Sciences Research Council. ML acknowledges the Agence
Nationale de la Recherche for its support, project number 06-BLAN-0067
\end{acknowledgements}

\bibliographystyle{aa} \bibliography{/home/lisbeth/tex/ref_lya}

\begin{appendix}

\section{Data for individual LAE candidates}

\subsection{Spectroscopically confirmed candidates}

This section gives the individual properties for the spectroscopically
confirmed candidates in each of the three survey fields. The
magnitudes in the tables are total magnitudes taken to be the
SExtractor MAG\_AUTO \citep{bertin1996}. From the total narrow-band
magnitude we compute Ly$\alpha$ flux, luminosity and star formation
rates. The EWs are computed based on colour indices computed from the
isophotal magnitudes.  The properties are derived as described in
detail in \cite{fynbo2002}. For objects where the measured flux was
below the $1\sigma$ level are indicated as lower/upper limits.

\begin{table*}
\begin{minipage}{\textwidth}
\caption{Properties of the 18 confirmed LAEs in the field of
BRI~1202--0725.}
\label{tab:lego_properties_1202}
\begin{tabular}{rrrrrrrrrrr}
\hline\hline
id & $\alpha$ (J2000) & $\delta$ (J2000) & z & N & $V$ & $R$ & EW$_0$ & f(Ly$\alpha$) & $L$(Ly$\alpha$) & SFR\\
 & & & & & & & {\AA} & $10^{-18}$ ergs s$^{-1}$ cm$^{-2}$ &  $10^{41}$ ergs s$^{-1}$ & M$_\odot$yr$^{-1}$\\

\hline

 2 & 12:05:14.0 & -07:40:05 & 3.2161 & $24.31^{+ 0.10}_{- 0.09}$ & $25.05^{+ 0.10}_{- 0.09}$ & $24.98^{+ 0.11}_{- 0.10}$ & $28^{+6}_{-5}$ & $  48.22^{+   3.96}_{-   4.73}$ & $  43.58^{+   3.58}_{-   4.28}$ &     4.47\\
\vspace{1mm}
 
 3 & 12:05:14.4 & -07:42:40 & 3.2012 & $22.19^{+ 0.01}_{- 0.01}$ & $23.43^{+ 0.02}_{- 0.02}$ & $23.63^{+ 0.03}_{- 0.03}$ & $28^{+1}_{-1}$ & $ 340.15^{+   4.56}_{-   4.68}$ & $ 307.46^{+   4.12}_{-   4.23}$ &    31.55\\
\vspace{1mm}
 
 4 & 12:05:14.5 & -07:40:10 & 3.1804 & $26.45^{+ 0.34}_{- 0.26}$ & $27.36^{+ 0.42}_{- 0.30}$ & $27.81^{+ 0.88}_{- 0.48}$ & $44^{+37}_{-19}$ & $    6.71^{+   1.41}_{-   2.45}$ & $   6.06^{+   1.28}_{-   2.21}$ &     0.62\\
\vspace{1mm}
 
 5 & 12:05:10.4 & -07:45.40 & 3.1821 & $23.25^{+ 0.03}_{- 0.03}$ & $24.62^{+ 0.05}_{- 0.05}$ & $25.07^{+ 0.09}_{- 0.08}$ & $34^{+2}_{-2}$ & $ 127.53^{+   3.12}_{-   3.28}$ & $ 115.27^{+   2.82}_{-   2.96}$ &    11.83\\
\vspace{1mm}
 
 7 & 12:05:27.4 & -07:40:38 & 3.2070 & $25.25^{+ 0.19}_{- 0.16}$ & $25.92^{+ 0.18}_{- 0.15}$ & $26.21^{+ 0.28}_{- 0.22}$ & $27^{+9}_{-7}$ & $  20.32^{+   2.76}_{-   3.80}$ & $  18.37^{+   2.50}_{-   3.43}$ &     1.89\\
\vspace{1mm}
 
 8\footnote{These objects were included in the same slit.} & 12:05:18.2 & -07:42:12 & 3.2106 & $25.67^{+ 0.23}_{- 0.19}$ & $26.93^{+ 0.41}_{- 0.30}$ & $\geq28.05$ & $64^{+44}_{-23}$ & $  13.73^{+   2.18}_{-   3.20}$ & $  12.41^{+   1.97}_{-   2.90}$ &     1.27\\
\vspace{1mm}
 
 9 & 12:05:25.3 & -07:41:13 & 3.2087 & $25.84^{+ 0.24}_{- 0.19}$ & $26.45^{+ 0.21}_{- 0.18}$ & $26.39^{+ 0.23}_{- 0.19}$ & $28^{+16}_{-11}$ & $  11.76^{+   1.93}_{-   2.88}$ & $  10.63^{+   1.75}_{-   2.60}$ &     1.09\\
\vspace{1mm}
 
10 & 12:05:23.7 & -07:43:44 & 3.2222 & $25.32^{+ 0.18}_{- 0.16}$ & $26.61^{+ 0.26}_{- 0.20}$ & $26.28^{+ 0.28}_{- 0.22}$ & $38^{+18}_{-12}$ & $  18.94^{+   2.60}_{-   3.59}$ & $  17.12^{+   2.35}_{-   3.24}$ &     1.76\\
\vspace{1mm}
 
11$^a$ & 12:05:18.2 & -07:42:09 & 3.2106 & $23.25^{+ 0.05}_{- 0.05}$ & $24.37^{+ 0.07}_{- 0.07}$ & $24.66^{+ 0.11}_{- 0.10}$ & $51^{+4}_{-4}$ & $  128.06^{+   5.57}_{-   6.10}$ & $ 115.75^{+   5.03}_{-   5.51}$ &    11.88\\
\vspace{1mm}
 
13 & 12:05:18.6 & -07:43:44 & 3.1992 & $24.81^{+ 0.07}_{- 0.07}$ & $26.77^{+ 0.25}_{- 0.20}$ & $27.57^{+ 0.74}_{- 0.44}$ & $74^{+19}_{-14}$ & $  30.44^{+   1.87}_{-   2.14}$ & $  27.51^{+   1.69}_{-   1.93}$ &     2.82\\
\vspace{1mm}
 
14 & 12:05:12.4 & -07:40:48 & 3.2191 & $24.91^{+ 0.12}_{- 0.11}$ & $\geq28.08$ & $\geq27.93$ & $82^{+29}_{-19}$ & $  27.82^{+   2.61}_{-   3.21}$ & $  25.15^{+   2.36}_{-   2.91}$ &     2.58\\
\vspace{1mm}
 
15 & 12:05:31.3 & -07:41:42 & 3.1862 & $25.84^{+ 0.19}_{- 0.16}$ & $27.68^{+ 0.65}_{- 0.41}$ & $28.12^{+ 1.63}_{- 0.62}$ & $75^{+62}_{-29}$ & $  11.72^{+   1.63}_{-   2.25}$ & $  10.59^{+   1.47}_{-   2.03}$ &     1.09\\
\vspace{1mm}
 
16 & 12:05:13.8 & -07:42:09 & 3.2029 & $25.80^{+ 0.22}_{- 0.19}$ & $\geq28.34$ & $\geq28.19$ & $106^{+107}_{-44}$ & $   12.22^{+   1.92}_{-   2.79}$ & $  11.05^{+   1.73}_{-   2.52}$ &     1.13\\
\vspace{1mm}
 
18 & 12:05:15.1 & -07:44:20 & 3.1926 & $25.24^{+ 0.17}_{- 0.15}$ & $\geq28.03$ & $\geq27.88$ & $103^{+75}_{-35}$ & $  20.51^{+   2.63}_{-   3.54}$ & $  18.54^{+   2.38}_{-   3.20}$ &     1.90\\
\vspace{1mm}
 
19 & 12:05:22.5 & -07:44:08 & 3.1963 & $25.56^{+ 0.16}_{- 0.14}$ & $\geq28.44$ & $\geq28.29$ & $\geq253$ & $    15.22^{+   1.83}_{-   2.41}$ & $  13.76^{+   1.65}_{-   2.17}$ &     1.41\\
\vspace{1mm}
 
21 & 12:05:18.6 & -07:43:44 & 3.1978 & $25.70^{+ 0.21}_{- 0.17}$ & $27.56^{+ 0.74}_{- 0.44}$ & $26.25^{+ 0.20}_{- 0.17}$ & $117^{+154}_{-50}$ & $  13.38^{+   1.97}_{-   2.79}$ & $  12.09^{+   1.78}_{-   2.52}$ &     1.24\\
\vspace{1mm}
 
22 & 12:05:19.1 & -07:42:31 & 3.2057 & $25.65^{+ 0.15}_{- 0.13}$ & $27.86^{+ 0.76}_{- 0.44}$ & $28.17^{+ 1.59}_{- 0.62}$ & $219^{+662}_{-104}$ & $  14.03^{+   1.58}_{-   2.05}$ & $  12.68^{+   1.43}_{-   1.85}$ &     1.30\\
\vspace{1mm}
 
23 & 12:05:24.1 & -07:44:01 & 3.2203 & $25.91^{+ 0.29}_{- 0.23}$ & $\geq28.22$ & $\geq28.07$ & $204^{+2165}_{-111}$ & $  10.98^{+   2.07}_{-   3.32}$ & $   9.93^{+   1.87}_{-   3.00}$ &     1.02\\

 \hline
\end{tabular}

\end{minipage}
\end{table*}

\begin{table*}
\caption{Properties of the 18 confirmed LAEs in the field of
BRI~1346--0322.}
\label{tab:lego_properties_1346}
\begin{tabular}{rrrrrrrrrrr}
\hline\hline
id & $\alpha$ (J2000) & $\delta$ (J2000) & z & N & $B$ & $R$ & EW$_0$ & f(Ly$\alpha$) & $L$(Ly$\alpha$) & SFR\\
 & & & & & & & {\AA} & $10^{-18}$ ergs s$^{-1}$ cm$^{-2}$ &  $10^{41}$ ergs s$^{-1}$ & M$_\odot$yr$^{-1}$ \\

\hline

 1 & 13:49:19.4 & -03:40:30 & 3.1592 & $24.28^{+ 0.12}_{- 0.11}$ & $26.60^{+ 0.46}_{- 0.32}$ & $\geq27.14$ & $462^{+1942}_{-240}$ & $  49.09^{+   4.73}_{-   5.86}$ & $  42.63^{+   4.11}_{-   5.09}$ &     4.37\\
\vspace{1mm}
 
 2 & 13:49:13.7 & -03:40:06 & 3.1307 & $24.52^{+ 0.13}_{- 0.11}$ & $\geq28.01$ & $26.97^{+ 1.27}_{- 0.57}$ & $327^{+2463}_{-164}$ & $  39.19^{+   3.85}_{-   4.80}$ & $  34.03^{+   3.35}_{-   4.17}$ &     3.49\\
\vspace{1mm}
 
 3 & 13:49:26.0 & -03:39:41 & 3.1602 & $25.62^{+ 0.23}_{- 0.19}$ & $\geq28.49$ & $\geq27.85$ & $95^{+173}_{-46}$ & $  14.28^{+   2.29}_{-   3.37}$ & $  12.40^{+   1.99}_{-   2.92}$ &     1.27\\
\vspace{1mm}
 
 4 & 13:49:27.7 & -03:39:40 & 3.1692 & $25.36^{+ 0.16}_{- 0.14}$ & $\geq28.60$ & $\geq27.96$ & $\geq290$ & $   18.14^{+   2.24}_{-   2.97}$ & $  15.75^{+   1.94}_{-   2.58}$ &     1.62\\
\vspace{1mm}
 
 5 & 13:49.15.7 & -03:39:37 & 3.1250 & $24.84^{+ 0.16}_{- 0.14}$ & $27.39^{+ 0.85}_{- 0.47}$ & $\geq27.44$ & $107^{+101}_{-40}$ & $  29.39^{+   3.49}_{-   4.58}$ & $  25.52^{+   3.03}_{-   3.98}$ &     2.62\\
\vspace{1mm}
 
 6 & 13:49:20.4 & -03:39:01 & 3.1345 & $24.82^{+ 0.14}_{- 0.13}$ & $\geq28.17$ & $25.55^{+ 0.20}_{- 0.17}$ & $\geq496$ & $  29.88^{+   3.29}_{-   4.23}$ & $  25.95^{+   2.86}_{-   3.67}$ &     2.66\\
\vspace{1mm}
 
 7 & 13:49:12.2 & -03:38:51 & 3.1804 & $24.16^{+ 0.10}_{- 0.09}$ & $25.40^{+ 0.13}_{- 0.12}$ & $24.69^{+ 0.11}_{- 0.10}$ & $22^{+5}_{-5}$ & $  54.64^{+   4.52}_{-   5.42}$ & $  47.45^{+   3.93}_{-   4.70}$ &     4.87\\
\vspace{1mm}
 
 8 & 13:49:16.4 & -03:38:32 & 3.1703 & $24.36^{+ 0.11}_{- 0.10}$ & $\geq27.96$ & $\geq27.32$ & $\geq585$ & $   45.68^{+   4.10}_{-   4.99}$ & $  39.66^{+   3.56}_{-   4.33}$ &     4.07\\
\vspace{1mm}
 
10 & 13:49:09.5 & -03:37:23 & 3.1559 & $24.97^{+ 0.17}_{- 0.15}$ & $26.97^{+ 0.49}_{- 0.34}$ & $26.91^{+ 0.99}_{- 0.51}$ & $64^{+55}_{-26}$ & $  26.08^{+   3.34}_{-   4.50}$ & $  22.65^{+   2.90}_{-   3.91}$ &     2.32\\ 
\vspace{1mm}
 
12 & 13:49:04.6 & -03:36:35 & 3.1843 & $25.28^{+ 0.15}_{- 0.13}$ & $26.53^{+ 0.19}_{- 0.16}$ & $25.70^{+ 0.14}_{- 0.13}$ & $35^{+10}_{-8}$ & $  19.48^{+   2.18}_{-   2.80}$ & $  16.92^{+   1.89}_{-   2.43}$ &     1.74\\
\vspace{1mm}
 
14 &13:49:05.7 & -03:35:03 & 3.1327 & $24.36^{+ 0.14}_{- 0.13}$ & $25.29^{+ 0.14}_{- 0.12}$ & $24.83^{+ 0.15}_{- 0.13}$ & $24^{+7}_{-6}$ & $  45.54^{+   4.98}_{-   6.37}$ & $  39.55^{+   4.32}_{-   5.53}$ &     4.06\\
\vspace{1mm}
 
17 & 13:49:14.5 & -03:35:26 & 3.1646 & $22.76^{+ 0.05}_{- 0.05}$ & $24.22^{+ 0.07}_{- 0.07}$ & $23.31^{+ 0.05}_{- 0.05}$ & $29^{+3}_{-3}$ & $199.39^{+   8.31}_{-   9.07}$ & $ 173.14^{+   7.22}_{-   7.87}$ &    17.77\\
\vspace{1mm}
 
20 & 13:49:20.0 & -03:35:52 & 3.1568 & $25.50^{+ 0.24}_{- 0.19}$ & $\geq28.33$ & $\geq27.70$ & $\geq264$ & $  16.00^{+   2.61}_{-   3.87}$ & $  13.90^{+   2.27}_{-   3.36}$ &     1.43\\
\vspace{1mm}
 
21 & 13:49:14.3 & -03:36:08 & 3.1729 & $24.29^{+ 0.13}_{- 0.11}$ & $\geq27.75$ & $25.47^{+ 0.27}_{- 0.22}$ & $168^{+434}_{-79}$ & $  48.57^{+   4.82}_{-   6.01}$ & $  42.18^{+   4.18}_{-   5.22}$ &     4.33\\
\vspace{1mm}
 
22 & 13:49:21.5 & -03:36:09 & 3.1371 & $25.64^{+ 0.20}_{- 0.17}$ & $27.09^{+ 0.31}_{- 0.24}$ & $24.61^{+ 0.05}_{- 0.05}$ & $43^{+15}_{-63}$ & $  14.06^{+   1.99}_{-   2.77}$ & $  12.21^{+   1.73}_{-   2.41}$ &     1.25\\
\vspace{1mm}
 
23 & 13:49:17.7 & -03:36:33 & 3.1468 & $24.63^{+ 0.13}_{- 0.12}$ & $\geq28.09$ &  $26.48^{+ 0.58}_{- 0.37}$ & $67^{+30}_{-19}$ & $  35.60^{+   3.58}_{-   4.48}$ & $  30.91^{+   3.11}_{-   3.89}$ &     3.17\\
\vspace{1mm}
 
24 & 13:49:06.7 & -03:36:32 & 3.1665 & $24.85^{+ 0.16}_{- 0.14}$ & $26.55^{+ 0.32}_{- 0.25}$ & $26.05^{+ 0.35}_{- 0.27}$ & $63^{+36}_{-20}$ & $ 29.17^{+   3.46}_{-   4.54}$ & $  25.33^{+   3.00}_{-   3.94}$ &     2.60\\
\vspace{1mm}
 
25 & 13:49:16.9 & -03:36:36 & 3.1717 & $25.54^{+ 0.20}_{- 0.17}$ & $\geq28.56$ & $\geq27.92$ & $129^{+478}_{-67}$ & $  15.41^{+   2.21}_{-   3.10}$ & $  13.38^{+   1.92}_{-   2.69}$ &     1.37\\
 
 \hline
\end{tabular}
\end{table*}

\begin{table*}
\begin{minipage}{\textwidth}
\caption{Properties of the 23 confirmed LAEs in the field of
Q~2138--4427.}
\label{tab:lego_properties_2138}
\begin{tabular}{rrrrrrrrrrrr}
\hline\hline
id & $\alpha$ (J2000) & $\delta$ (J2000) & z & N & $B$ & $R$ & EW$_0$ & f(Ly$\alpha$) & $L$(Ly$\alpha$) & SFR\\
 & & & & & & & {\AA} & $10^{-18}$ ergs s$^{-1}$ cm$^{-2}$ &  $10^{41}$ ergs s$^{-1}$ & M$_\odot$yr$^{-1}$\\
\hline

 4 & 21:42:12.6 & -44:15:34 & 2.8525 & $25.15^{+ 0.11}_{- 0.10}$ & $27.21^{+ 0.51}_{- 0.34}$ & $25.53^{+ 0.17}_{- 0.15}$ & $75^{+44}_{-24}$ & $  28.55^{+   2.49}_{-   3.01}$ & $  19.47^{+   1.69}_{-   2.05}$ &     2.00\\
\vspace{1mm}
 
 8 & 21:41:56.2 & -44:15:10 & 2.8528 & $26.97^{+ 0.29}_{- 0.23}$ & $28.04^{+ 0.55}_{- 0.36}$ & $26.30^{+ 0.17}_{- 0.15}$ & $33^{+46}_{-18}$ &$   5.37^{+   1.01}_{-   1.63}$ & $   3.66^{+   0.69}_{-   1.11}$ &     0.38\\
\vspace{1mm}
 
10 & 21:41:54.4 & -44:14:22 & 2.8487 & $25.68^{+ 0.18}_{- 0.15}$ & $\geq28.30$ & $\geq27.63$ & $\geq440$ & $  17.66^{+   2.34}_{-   3.18}$ & $  12.04^{+   1.59}_{-   2.17}$ &     1.24\\
\vspace{1mm}
 
11 & 21:42:03.7 & -44:14:16 & 2.8563 & $25.11^{+ 0.13}_{- 0.11}$ & $25.80^{+ 0.16}_{- 0.14}$ & $24.06^{+ 0.06}_{- 0.05}$ & $38^{+21}_{-13}$ &  $  29.77^{+   2.96}_{-   3.70}$ & $  20.30^{+   2.02}_{-   2.52}$ &     2.08\\
\vspace{1mm}
 
12 & 21:42:01.9 & -44:13:52 & 2.8576 & $25.17^{+ 0.14}_{- 0.12}$ & $26.76^{+ 0.40}_{- 0.29}$ & $27.02^{+ 1.28}_{- 0.57}$  & $32^{+18}_{-11}$ &  $  28.00^{+   2.94}_{-   3.73}$ & $  19.09^{+   2.01}_{-   2.54}$ &     1.96\\
\vspace{1mm}
 
14 & 21.41:49.0 & -44:13:46 & 2.8463 & $24.23^{+ 0.07}_{- 0.06}$ & $25.33^{+ 0.12}_{- 0.11}$ & $25.99^{+ 0.42}_{- 0.30}$ & $49^{+12}_{-9}$ & $  66.75^{+   3.84}_{-   4.34}$ & $  45.52^{+   2.62}_{-   2.96}$ &     4.67\\
\vspace{1mm}
 
16 &  21:41:58.9 & -44:10:10 & 2.8561 &  $25.64^{+ 0.15}_{- 0.13}$ & $26.57^{+ 0.24}_{- 0.20}$ & $26.43^{+ 0.38}_{- 0.28}$  & $25^{+13}_{-9}$ & $18.32^{+   2.12}_{-   2.75}$ & $  12.49^{+   1.44}_{-   1.88}$ &     1.28\\
\vspace{1mm}
 
17 & 21:42:15.5 & -44:11:40 & 2.8532 &  $24.73^{+ 0.07}_{- 0.07}$ &  $27.17^{+ 0.46}_{- 0.32}$ & $25.56^{+ 0.17}_{- 0.14}$ & $218^{+197}_{-74}$ & $  42.13^{+   2.58}_{-   2.95}$ & $  28.73^{+   1.76}_{-   2.01}$ &     2.95\\
\vspace{1mm}
 
18 & 21:41:44.7 & -44:11:23 & 2.8627 & $24.56^{+ 0.07}_{- 0.07}$ & $27.88^{+ 1.68}_{- 0.63}$ & $25.24^{+ 0.15}_{- 0.13}$ & $253^{+266}_{-90}$ &  $  49.28^{+   3.07}_{-   3.51}$ & $  33.60^{+   2.09}_{-   2.39}$ &     3.45\\
\vspace{1mm}
 
19 & 21:42:14.9 & -44:10:45 & 2.8542 & $25.84^{+ 0.17}_{- 0.15}$ &  $27.44^{+ 0.55}_{- 0.36}$ & $\geq27.84$ & $30^{+20}_{-12}$ & $  15.19^{+   1.96}_{-   2.64}$ & $  10.36^{+   1.33}_{-   1.80}$ &     1.06\\
\vspace{1mm}
 
20 & 21:41:58.0 & -44:10:41 & 2.8564 &  $26.34^{+ 0.21}_{- 0.18}$ & $\geq28.82$ & $\geq27.84$  & $\geq243$ & $   9.58^{+   1.45}_{-   2.07}$ & $   6.53^{+   0.99}_{-   1.41}$ &     0.67\\
\vspace{1mm}
 
21 & 21:41:44.0 & -44:10:59 & 2.8525 & $26.54^{+ 0.20}_{- 0.17}$ & $\geq29.06$ & $\geq28.42$ & $\geq437$ & $   7.99^{+   1.17}_{-   1.66}$ & $   5.45^{+   0.80}_{-   1.13}$ &     0.56\\
\vspace{1mm}
 
22 & 21:42:11.1 & -44:11:05 & 2.8537 &  $26.02^{+ 0.17}_{- 0.15}$ & $\geq28.72$ & $\geq28.06$  & $264^{+2331}_{-167}$ & $  12.80^{+   1.65}_{-   2.22}$ & $   8.73^{+   1.12}_{-   1.51}$ &     0.90 \\
\vspace{1mm}
 
23 & 21:41:51.4 & -44:11:03 & 2.8593 & $25.77^{+ 0.16}_{- 0.14}$ & $27.67^{+ 0.72}_{- 0.43}$ & $27.43^{+ 1.25}_{- 0.57}$ & $92^{+140}_{-41}$ & $  16.18^{+   1.98}_{-   2.62}$ & $  11.03^{+   1.35}_{-   1.79}$ &     1.13 \\
\vspace{1mm}
 
25 & 21:41:49.1 & -44:11:13 & 2.8608 &  $24.94^{+ 0.11}_{- 0.10}$ &  $27.02^{+ 0.53}_{- 0.36}$ & $\geq27.43$ & $40^{+18}_{-12}$ & $  34.87^{+   3.03}_{-   3.67}$ & $  23.78^{+   2.07}_{-   2.50}$ &     2.44\\
\vspace{1mm}
 
26 & 21:41:43.3 & -44:11:25 & 2.8617 & $24.61^{+ 0.08}_{- 0.08}$ & $27.93^{+ 2.33}_{- 0.69}$ & $25.98^{+ 0.34}_{- 0.26}$ & $86^{+37}_{-22}$ & $  46.95^{+   3.20}_{-   3.71}$ & $  32.01^{+   2.18}_{-   2.53}$ &     3.29\\
\vspace{1mm}
 
27 & 21:41:48.1 & -44:11:19 & 2.8594 & $25.68^{+ 0.16}_{- 0.14}$ & $\geq28.43$ & $\geq27.77$ & $167^{+1164}_{-88}$ & $  17.59^{+   2.12}_{-   2.79}$ & $  11.99^{+   1.44}_{-   1.90}$ &     1.23\\
\vspace{1mm}
 
29 & 21:41:49.2 & -44:11:48 & 2.8607 & $23.19^{+ 0.03}_{- 0.03}$ & $24.48^{+ 0.06}_{- 0.06}$ & $24.04^{+ 0.06}_{- 0.06}$ & $31^{+3}_{-3}$ & $ 174.44^{+   4.63}_{-   4.88}$ & $ 118.96^{+   3.15}_{-   3.33}$ &    12.21 \\
\vspace{1mm}
 
30 & 21:41:59.2 & -44:11:53 & 2.8547 &  $26.02^{+ 0.18}_{- 0.15}$ & $\geq28.68$ & $\geq28.04$    & $234^{+2360}_{-145}$ & $  12.84^{+   1.67}_{-   2.26}$ & $   8.76^{+   1.14}_{-   1.54}$ &     0.90\\ 
\vspace{1mm}
 
31 & 21:42:08.1 & -44:12:01 & 2.8652 &  $26.56^{+ 0.26}_{- 0.21}$ & $27.27^{+ 0.33}_{- 0.25}$ & $26.64^{+ 0.31}_{- 0.24}$ & $22^{+21}_{-12}$ & $   7.79^{+   1.35}_{-   2.08}$ & $   5.31^{+   0.92}_{-   1.42}$ &     0.54\\
\vspace{1mm}
 
32 &  21:41:49.1 & -44:13:05 & 2.8779 & $24.91^{+ 0.09}_{- 0.08}$ & $25.93^{+ 0.15}_{- 0.13}$ & $26.04^{+ 0.29}_{- 0.23}$ & $20^{+6}_{-5}$ & $  35.80^{+   2.64}_{-   3.10}$ & $  24.41^{+   1.80}_{-   2.11}$ &     2.51\\
\vspace{1mm}
 
33 & 21:41:47.5 & -44:13:14 & 2.8591 & $25.01^{+ 0.12}_{- 0.11}$ & $27.25^{+ 0.74}_{- 0.44}$ & $25.92^{+ 0.33}_{- 0.25}$  & $22^{+9}_{-6}$ & $  32.71^{+   3.09}_{-   3.81}$ & $  22.30^{+   2.11}_{-   2.60}$ &     2.29\\
\vspace{1mm}
 
36\footnote{This object was identified visually as a blend with
another much brighter source. Therefore it has not been able to
measure the photometric properties of this candidate.} & 21:41:59.3 &
-44:13:18 & 2.8586 \\

\hline
\end{tabular}
\end{minipage}
\end{table*}

\subsection{Candidates not confirmed by the available spectroscopy}

This section gives the properties of the emission line candidates for
which the nature could not be assessed through the available
spectroscopic data. The properties are measured as detailed in the
previous section. The restframe EWs are given assuming the redshift
corresponding to the central wavelength of the filter.

\begin{table*}
\caption{Properties of the non-confirmed LAE candidates.}
\label{tab:nonconf_prop}
\begin{minipage}{\textwidth}
\begin{center}
\begin{tabular}{rrrrrrrrrr}
\hline\hline
id & $\alpha$ (J2000) & $\delta$ (J2000) & N & $V$/$B$ & $R$ & EW$_0$ & f(Ly$\alpha$) & $L$(Ly$\alpha$) & SFR\\
 & & & & & & {\AA} & $10^{-18}$ ergs\; s$^{-1}$ cm$^{-2}$ &  $10^{41}$ ergs\; s$^{-1}$ \\

\hline
BRI~1202--0725 \\
\hline
 1 & 12:05:31.5 & -07:41:16 & $26.05^{+ 0.29}_{- 0.23}$ & $26.53^{+ 0.23}_{- 0.19}$ & $26.26^{+ 0.20}_{- 0.17}$ & $21^{+13}_{-9}$ & $   9.73^{+   1.84}_{-   2.96}$ & $   8.79^{+   1.66}_{-   2.67}$ &     0.90\\
\vspace{1mm}
 
 6 & 12:05:31.2 & -07:42:48 & $26.42^{+ 0.36}_{- 0.27}$ & $26.99^{+ 0.30}_{- 0.24}$ & $27.30^{+ 0.51}_{- 0.34}$ & $27^{+19}_{-12}$ & $   6.87^{+   1.50}_{-   2.68}$ & $   6.21^{+   1.36}_{-   2.42}$ &     0.64\\
\vspace{1mm}
 
12 & 12:05:27.7 & -07:42:38 & $26.55^{+ 0.37}_{- 0.28}$ & $26.86^{+ 0.24}_{- 0.20}$ & $26.47^{+ 0.19}_{- 0.16}$ & $42^{+34}_{-18}$ & $   6.14^{+   1.38}_{-   2.50}$ & $   5.55^{+   1.25}_{-   2.26}$ &     0.57\\
\vspace{1mm}
 
17 & 12:05:10.6 & -07:41:06 & $26.56^{+ 0.22}_{- 0.18}$ & $27.89^{+ 0.42}_{- 0.30}$ & $\geq28.98$ & $71^{+107}_{-35}$ & $   6.04^{+   0.93}_{-   1.34}$ & $   5.46^{+   0.84}_{-   1.21}$ &     0.56\\
\vspace{1mm}
 
20 & 12:05:31.9 & -07:41:30 & $26.85^{+ 0.22}_{- 0.18}$ & $\geq29.42$ & $28.63^{+ 0.87}_{- 0.48}$ & $186^{+2195}_{-109}$ & $   4.62^{+   0.71}_{-   1.02}$ & $4.17^{+   0.64}_{-   0.93}$ &     0.43\\
\vspace{1mm}
 
24 & 12:05:21.0 & -07:45:01 &  $26.97^{+ 0.50}_{- 0.34}$ & $27.06^{+ 0.25}_{- 0.21}$ & $26.76^{+ 0.22}_{- 0.18}$ & $74^{+70}_{-32}$ &  $4.17^{+   1.13}_{-   2.45}$ & $   3.76^{+   1.02}_{-   2.22}$ & $0.39$\\
\vspace{1mm}
 
25 & 12:05:28.3 & -07:40:54 & $26.88^{+ 0.36}_{- 0.27}$ & $28.29^{+ 0.81}_{- 0.46}$ & $27.46^{+ 0.36}_{- 0.27}$ & $118^{+370}_{-61}$ & $   4.51^{+   0.98}_{-   1.74}$ & $   4.07^{+   0.89}_{-   1.58}$ &     0.42\\
\hline
BRI~1346--0322\\
\hline 
 9 & 13:49:10.5 & -03:38:21 & $25.02^{+ 0.17}_{- 0.15}$ & $26.42^{+ 0.26}_{- 0.21}$ & $24.39^{+ 0.07}_{- 0.06}$ & $28^{+11}_{-8}$ & $  24.86^{+   3.16}_{-   4.24}$ & $  21.59^{+   2.75}_{-   3.68}$ &     2.22\\
\vspace{1mm}
 
11\footnote{These objects have not been observed spectroscopically.} & 13:49:27.3 & -03:33:58 & $25.76^{+ 0.24}_{- 0.20}$ &  $27.43^{+ 0.50}_{- 0.34}$ & $\geq27.94$ & $62^{+62}_{-27}$ & $  12.57^{+   2.10}_{-   3.16}$ & $  10.92^{+   1.83}_{-   2.74}$ &     1.12\\
\vspace{1mm}
 
13 & 13:49:26.4 & -03:34:02 & $25.86^{+ 0.28}_{- 0.22}$ & $26.90^{+ 0.30}_{- 0.24}$ & $25.98^{+ 0.21}_{- 0.18}$ & $93^{+130}_{-41}$ & $  11.42^{+   2.12}_{-   3.36}$ & $   9.92^{+   1.84}_{-   2.92}$ &     1.02\\
\vspace{1mm}
 
15$^a$ & 13:49:29.4 & -03:35:08 & $25.36^{+ 0.23}_{- 0.19}$ & $26.29^{+ 0.22}_{- 0.18}$ & $24.60^{+ 0.08}_{- 0.07}$ & $32^{+18}_{-12}$ & $  18.08^{+   2.89}_{-   4.24}$ & $  15.70^{+   2.51}_{-   3.68}$ &     1.61\\
\vspace{1mm}
 
18$^a$ & 13:49:29.2 & -03:35:31.2 & $25.23^{+ 0.20}_{- 0.17}$ & $26.53^{+ 0.28}_{- 0.22}$ & $24.57^{+ 0.07}_{- 0.07}$ & $34^{+22}_{-13}$ & $  20.51^{+   2.98}_{-   4.19}$ & $  17.81^{+   2.58}_{-   3.64}$ &     1.83\\
\vspace{1mm}
 
26 & 13:49:19.4 & -03:37:01 & $25.60^{+ 0.17}_{- 0.14}$ & $27.14^{+ 0.29}_{- 0.23}$ & $25.20^{+ 0.08}_{- 0.07}$ & $26^{+13}_{-9}$ & $  14.61^{+   1.81}_{-   2.41}$ & $  12.68^{+   1.57}_{-   2.09}$ &     1.30\\
\hline
Q~2138--4427\\
\hline
 1 & 21:41:44.6 & -44:16:07 &  $26.61^{+ 0.23}_{- 0.19}$ &  $28.34^{+ 0.91}_{- 0.49}$ & $25.44^{+ 0.09}_{- 0.08}$  & $31^{+31}_{-15}$ & $   7.45^{+   1.20}_{-   1.77}$ & $   5.08^{+   0.82}_{-   1.21}$ &     0.52\\
\vspace{1mm}
 
 2 & 21:41:53.0 & -44:15:42 &  $25.31^{+ 0.10}_{- 0.09}$ & $26.15^{+ 0.14}_{- 0.12}$ & $24.71^{+ 0.06}_{- 0.06}$ & $18^{+6}_{-5}$ & $  24.73^{+   1.93}_{-   2.29}$ & $  16.86^{+   1.32}_{-   1.56}$ &     1.73 \\
\vspace{1mm}
 
 3 & 21:41:46.2 & -44:15:38&  $25.55^{+ 0.16}_{- 0.14}$ & $27.06^{+ 0.44}_{- 0.31}$ & $25.32^{+ 0.14}_{- 0.12}$ & $35^{+19}_{-12}$ & $  19.75^{+   2.35}_{-   3.09}$ & $  13.47^{+   1.60}_{-   2.11}$ &     1.38\\
\vspace{1mm}
 
 5 & 21:42:12.1 & -44:15:32 &  $26.71^{+ 0.21}_{- 0.17}$ & $27.59^{+ 0.32}_{- 0.25}$ & $26.41^{+ 0.17}_{- 0.15}$& $23^{+19}_{-11}$ &  $   6.79^{+   1.01}_{-   1.43}$ & $   4.63^{+   0.69}_{-   0.98}$ &     0.48\\
\vspace{1mm}
 
 6 & 21:42:14.8 & -44:15:15 &  $26.72^{+ 0.26}_{- 0.21}$ & $27.49^{+ 0.35}_{- 0.27}$ & $26.08^{+ 0.15}_{- 0.14}$ & $29^{+38}_{-17}$ &$   6.72^{+   1.17}_{-   1.80}$ & $   4.58^{+   0.80}_{-   1.23}$ &     0.47\\
\vspace{1mm}
 
 7 & 21:42:12.6 & -44:15:11 &$26.31^{+ 0.22}_{- 0.18}$ & $27.50^{+ 0.45}_{- 0.32}$ & $25.57^{+ 0.12}_{- 0.11}$ & $29^{+27}_{-14}$ &$   9.82^{+   1.51}_{-   2.19}$ & $   6.69^{+   1.03}_{-   1.49}$ &     0.69\\
\vspace{1mm}
 
 9 & 21:42:10.5 & -44:14:38 & $26.33^{+ 0.17}_{- 0.15}$ & $27.66^{+ 0.40}_{- 0.29}$ & $25.72^{+ 0.11}_{- 0.10}$ & $50^{+59}_{-23}$ & $   9.62^{+   1.24}_{-   1.67}$ & $   6.56^{+   0.85}_{-   1.14}$ &     0.67\\
\vspace{1mm}
 
13 & 21:42:15.5 & -44:13:51 &  $26.41^{+ 0.17}_{- 0.15}$ & $27.21^{+ 0.24}_{- 0.20}$ & $26.16^{+ 0.15}_{- 0.13}$ & $20^{+13}_{-9}$ &  $   9.00^{+   1.15}_{-   1.54}$ & $   6.14^{+   0.78}_{-   1.05}$ &     0.63\\
\vspace{1mm}
 
15 & 21:42:12.9 & -44:13:35 & $25.97^{+ 0.18}_{- 0.16}$ & $27.43^{+ 0.50}_{- 0.34}$ & $26.03^{+ 0.22}_{- 0.18}$& $35^{+48}_{-19}$ &  $  13.41^{+   1.81}_{-   2.49}$ & $   9.15^{+   1.24}_{-   1.69}$ &     0.94 \\
\vspace{1mm}
 
34 & 21:42:12.6 & -44:11:36 &  $26.01^{+ 0.22}_{- 0.18}$ & $27.08^{+ 0.41}_{- 0.30}$ & $25.71^{+ 0.19}_{- 0.16}$ & $24^{+30}_{-14}$ &$  13.02^{+   2.04}_{-   2.97}$ & $   8.88^{+   1.39}_{-   2.02}$ &     0.91 \\
\vspace{1mm}
 
35 & 21:42:12.6 & -44:10:10 &$26.52^{+ 0.23}_{- 0.19}$ &  $28.09^{+ 0.75}_{- 0.44}$ & $26.26^{+ 0.20}_{- 0.17}$ & $41^{+50}_{-20}$ &$   8.08^{+   1.31}_{-   1.94}$ & $   5.51^{+   0.89}_{-   1.32}$ &     0.57\\

\hline
\end{tabular}
\end{center}
\end{minipage}
\end{table*}

\end{appendix}

\end{document}